\documentclass[12pt,preprint]{aastex}
\shorttitle{Fitting YSO SEDs using pre-computed model SEDs}
\shortauthors{Robitaille et al.}

\newcommand{\mstar}{M_{\star}}
\newcommand{\lstar}{L_{\rm bol}}

\newcommand{\mdisk}{M_{\rm{disk}}}

\newcommand{\agestar}{t_{\star}}
\newcommand{\tstar}{T_{\star}}
\newcommand{\rstar}{R_{\star}}
\newcommand{\rsub}{R_{\rm sub}}

\newcommand{\msol}{M_{\odot}}
\newcommand{\rsol}{R_{\odot}}
\newcommand{\lsol}{L_{\odot}}

\newcommand{\mipsns}{MIPS~24$\mu$m}
\newcommand{\av}{A_{\rm V}}
\newcommand{\rmind}{R^{\rm disk}_{\rm min}}
\newcommand{\rmaxd}{R^{\rm disk}_{\rm max}}
\newcommand{\mdote}{\dot{M}_{\rm env}}

\newcommand{\rnu}{R_{\nu}}
\newcommand{\fnua}{F_{\nu}{\rm{[actual]}}}
\newcommand{\fnuq}{F_{\nu}{\rm{[assumed]}}}
\newcommand{\fnuqz}{F_{\nu_0}{\rm{[quoted]}}}
\newcommand{\fnuqiso}{F_{\nu_0}^{\rm iso}{\rm{[quoted]}}}

\newcommand{\microns}{\,$\mu$m~}
\newcommand{\micronsns}{\,$\mu$m}

\newcommand{\rv}{R_{\rm V}}

\begin{document}

%%%%%%%%%%%%%%%%% Title Section %%%%%%%%%%%%%%%%%

\title{Interpreting Spectral Energy Distributions from Young Stellar Objects. II. Fitting observed SEDs using a large grid of pre-computed models}

\author{Thomas P. Robitaille\altaffilmark{1}}
\author{Barbara A. Whitney\altaffilmark{2}}
\author{Remy Indebetouw\altaffilmark{3}}
\author{Kenneth Wood\altaffilmark{1}}

\altaffiltext{1}{SUPA, School of Physics and Astronomy, University of St Andrews, North 
Haugh, KY16 9SS, St Andrews, United Kingdom; tr9@st-andrews.ac.uk, kw25@st-andrews.ac.uk}
\altaffiltext{2}{Space Science Institute, 4750 Walnut St. Suite 205, Boulder, CO 80301, USA; bwhitney@spacescience.org}
\altaffiltext{3}{Spitzer Fellow, University of Virginia, Astronomy Dept., P.O. Box 3818, Charlottesville, VA, 22903-0818; remy@virginia.edu}

%%%%%%%%%%%%%%%%% Abstract %%%%%%%%%%%%%%%%%

\begin{abstract}
We present a method to analyze the spectral energy distributions (SEDs) of young stellar objects (YSOs).
Our approach is to fit data with pre-computed 2-D radiation transfer models spanning a large region of parameter space.
This allows us to determine not only a single set of physical parameter values but the entire range of values consistent with the multi-wavelength observations of a given source.
In this way we hope to avoid any over-interpretation when modeling a set of data.
We have constructed spectral energy distributions from optical to sub-mm wavelengths, including new {\it Spitzer} IRAC and MIPS photometry, for 30 young and spatially resolved sources in the Taurus-Auriga star-forming region.
We demonstrate fitting model SEDs to these sources, and find that we correctly identify the evolutionary stage and physical parameters found from previous independent studies, such as disk mass, disk accretion rate, and stellar temperature.
We also explore how fluxes at various wavelengths help to constrain physical parameters, and show examples of degeneracies that can occur when fitting SEDs.
A web-based version of this tool is available to the community\footnote{http://www.astro.wisc.edu/protostars}.
\end{abstract}


\section{Introduction}

\label{s:intro}

Studying the circumstellar environment of YSOs is essential to our understanding of the pre-main-sequence evolution of stars.
Since it is not possible to observe a single YSO through various stages of evolution over thousands or millions of years, this has to be done statistically, by observing large numbers of YSOs, and inferring from these possible evolutionary sequences.
Observations of resolved YSOs have been made over the past decade \citep[e.g.][]{tamura91,kenyon93p2,burrows96,whitney97,close97,lucas97,stapelfeldt98,padgett99,cotera01}, giving us some insight into the evolution of low-mass YSOs in nearby star-forming regions.
However, only a limited number of YSOs can be directly resolved ($\sim100$), most of which only marginally.
To obtain a quantitative picture of both low and high-mass star-formation, we require observations of a larger sample, by studying the tens of thousands of unresolved YSOs seen in nearby star-forming regions (e.g Taurus-Auriga, Perseus, Orion) as well as more distant star-forming regions (e.g. M16, M17, NGC6334).

In order to study unresolved or close-to unresolved YSOs in these distant regions, we can resort to multi-wavelength photometry, from which we construct SEDs.
The main question we seek to answer in this series of papers is: how can we make the most of the information contained in the SEDs while limiting any over-interpretation?

Many methods have been suggested and used in the past to classify and interpret YSO SEDs, such as spectral indices \citep[e.g.][]{lada87} or color-color diagrams \citep[e.g.][]{lada92,allen04}.
In \citet[][hereafter Paper I]{robitaille06}, we presented a grid of 200,000 model SEDs spanning a large range of possible evolutionary stages and stellar masses, and using these models, we explored what could be learned from using spectral indices and color-color plots.
The main advantage of these two techniques is that they make it easy and fast to classify the SEDs of sources, although it is not necessarily trivial to extract information relating to the physical conditions in these objects.
Furthermore, many YSOs now have data at more than four wavelengths, and although color-color or color-magnitude plots with more than two dimensions can be constructed to make use of this extra information, only sources that have data in all of the required filters can be used.
In large-scale multi-wavelength studies of star-formation, it is not uncommon to have a significant number of sources that lack at least one datapoint.
Therefore, it is usually not possible to make the best use of all of the data available for each source using these techniques.

Another approach to analyzing SEDs is to compute radiation transfer models, first assuming a given circumstellar dust and gas geometry, as well as dust properties, predicting the emergent SED, and finding a set of parameters that reproduce the observations - it is a typical \emph{inverse problem}.
The main advantage of this technique is that one gets an insight into the actual physical properties of the source, rather than simply quantifying the shape of the SED.
Another advantage of using SED fitting is that it can make use of any data available, while not being limited by the lack of a datapoint at a given wavelength.
Of course, the more data are present, the better the parameters will be constrained.

If a source has been observed at many different wavelengths from optical to mm wavelengths, it is probable, though not certain, that the set of parameters for the model SED that will match the data well will be fairly unique, provided that all the parameters actually affect the shape of the SED, and that there are no degeneracies.
However, many star-forming regions have been covered by several surveys, providing flux measurements at only half a dozen to a dozen wavelengths (e.g. 2MASS, {\it Spitzer} IRAC \& MIPS; \citealt{skrutskie06, fazio04, rieke04, werner04}) for large numbers of sources.
When studying a large number of YSOs, one wants to know not only \emph{one} set of physical parameters that can explain the data for each source, but \emph{all} the different sets of parameters that can explain the data for each source, in order to avoid any over-interpretation.
Fitting SEDs to these many sources by trial and error would be problematic: it is very likely that there is not a unique set of parameters that can explain the data available for a given source, and it would be extremely time-consuming to explore parameter space manually for each source, let alone thousands of sources.

Our proposed solution to this problem is to precompute a large set of radiation transfer models that provide a reasonably large coverage of parameter space.
We can then use this set of models to find all the combinations of parameters that can explain the data for a given source.

This paper describes the method that we use to fit these models to data for single sources, and presents examples of what can be learned from using such a technique.
In Section~\ref{s:tech} we describe the technical details of the fitting method.
In Section~\ref{s:results} we demonstrate the use of the fitting method on Taurus-Auriga sources as a proof-of-concept: we first describe the sample of sources and the data used to construct the SEDs (\S\ref{s:data}); then we compare the values of a selection of physical parameters with independent estimates (\S\ref{s:poc}); and finally we show how different combinations of datapoints constrain physical parameters for a given source, and discuss degeneracies that can arise (\S\ref{s:degen}).
Additionally, we show an example of a resolved source for which we can measure fluxes in apertures smaller than the source itself, and demonstrate that we can fit the SEDs from the different apertures simultaneously (\S\ref{s:resolved}).
An example of large-scale study of a star-forming region will be presented in Robitaille et al. (2007, in preparation) and in Indebetouw et al. (2007, in preparation).

\section{Technical overview}

\label{s:tech}

The SED fitting tool that we have developed uses the 200,000 YSO model SEDs (20,000 sets of physical parameters and 10 viewing angles) presented in Paper I.
The models consist of pre-main-sequence stars with different combinations of axisymmetric circumstellar disks, infalling flattened envelopes, and outflow cavities, covering a wide range of parameter space.
The main caveat for this work is that we assume that stars of all masses form via accretion through a disk and envelope.
However, we note that the grid of models covers a large range of parameter space, so that we assume as little as possible about the specifics of the accretion physics.

The parameter ranges covered by the models span those determined from observations and theories.
The parameters we vary fall into three categories: the central source parameters (stellar mass, radius and temperature), the infalling envelope parameters (the envelope accretion rate, outer radius, inner radius, cavity opening angle and cavity density), and the disk parameters (disk mass, accretion rate, outer radius, inner radius, flaring power, and scaleheight).
The stellar masses $\mstar$ are sampled between $0.1$ and $50$\,$\msol$, the
stellar ages $\agestar$ are sampled between $10^3$ and $10^7$\,yr, and the
stellar radii $\rstar$ and temperatures $\tstar$ are found from $\mstar$ and
$\agestar$ using evolutionary tracks \citep{bernasconi96,siess00}.
The disk and envelope parameters are then sampled randomly within ranges that depend on the age of the central source.
For example, the disk mass was sampled from $\mdisk/\mstar\sim0.001-0.1$ at early evolutionary stages, and a wider range of masses extending down to $\mdisk/\mstar=10^{-8}$ between $1$ and $10$\,Myr to allow for the disk dispersal stage.
The dust grain models used in the radiation transfer models are the following: the densest regions of the disk ($n_{\rm H_2} > 10^{10}$ cm$^{-3}$) use a grain model with a size distribution that decays exponentially for sizes larger than 50\microns extending up to $1$\,mm, and the rest of the circumstellar geometry uses a grain size distribution with an average particle size slightly larger than the diffuse ISM, and a ratio of total-to-selective extinction $\rv=3.6$.
We assume a gas-to-dust ratio of 100 (note that the results can be scaled to different gas-to-dust ratios since only the dust is taken into account in the radiation transfer).
For more details about the models and the ranges of parameters sampled, including caveats, we refer readers to Paper I.

We have convolved these models with common filter bandpasses ranging from optical (e.g. UBVRI), to near and mid-IR (e.g. 2MASS JHK, {\it Spitzer} IRAC \& MIPS), far-IR, and sub-mm (e.g. IRAS, SCUBA).
The details of the convolution of the model SEDs with the filters used in this paper is described in Appendix~\ref{a:conv}.
A large range of filters is available in the online fitting tool, and additional filters can be added on request from users.

Let us assume that we have photometry of a source at N wavelengths $\lambda_{i~(i=1...{\rm N})}$, and that the flux densities are $F_{\nu}(\lambda_i)$ with uncertainties $\sigma(\lambda_i)$. In addition, we assume that the fluxes were measured in apertures $A(\lambda_i)$. Finally, we assume that the source lies in a distance range $d_{\rm min}$ to $d_{\rm max}$.

In order to fit convolved model fluxes to data we first interpolate the fluxes to the apertures $A(\lambda_i)$ used to perform the photometry, for a number of distances $d$ between $d_{\rm min}$ and $d_{\rm max}$, and scale them to the appropriate distance.
We are able to do this since each SED was computed for 50 apertures between 100 and 100,000\,AU.

At each of these distances, we fit the convolved model fluxes to the data using optimal scaling, allowing the visual extinction $\av$ to be a free parameter. 
Writing the convolved, interpolated, and scaled model fluxes as $M_{\nu}(\lambda_i)$, the overall pattern $P_{\nu}(\lambda_i)$ that is being fit to the data is then

\begin{equation}
P_{\nu}(\lambda_i)= M_{\nu}(\lambda_i)\times10^{-0.4\av \kappa_{\lambda_i}/\kappa_{\rm V}}
\end{equation}

where $\kappa_{\lambda_i}$ and $\kappa_{\rm V}$ are given by an extinction law. We use an extinction model calculated with the method of \citet*{kmh94} that fits a typical Galactic ISM curve modified for the mid-IR extinction properties derived by \cite{indebetouw05} (Wolff, private communication). Since $P_{\nu}(\lambda_i)$ is nonlinear, it is convenient to work with $\log{[F_{\nu}]}$ and $\log{[P_{\nu}]}$ rather than $F_{\nu}$ and $P_{\nu}$ so as to transform this into a linear problem. The model that is fit to the data is then:

\begin{equation}
\label{e:lin}
\log_{10}{[P_{\nu}(\lambda_i)]}=\log_{10}{[M_{\nu}(\lambda_i)]}-0.4\av\frac{\kappa_{\lambda_i}}{\kappa_{\rm V}}
\end{equation}

where the free parameter $\av$ can be determined from optimal scaling. The unbiased fluxes and flux variance in $\log_{10}$ space are given by:

\begin{eqnarray}
\langle\,\log_{10}{[F_{\nu}]}\,\rangle&=&\log_{10}\langle F_{\nu}\rangle\\&&-\frac{1}{2}\frac{1}{\log_{e}{10}}\frac{1}{\langle \,F_{\nu}\,\rangle^{2}}\,\sigma^{2}(F_{\nu})+...\\
\sigma^{2}(\langle\,\log_{10}{[F_{\nu}]}\,\rangle)&=&\left[\frac{1}{\log_{e}10}\frac{1}{\langle\,F_{\nu}\,\rangle}\right]^{2}\sigma^{2}(F_{\nu})+...
\end{eqnarray}

Once the free parameter $\av$ has been determined, we compute the $\chi^2$-per-datapoint value of the fit:

\begin{equation}
\chi^2=\frac{1}{N}\sum_{i=1}^N \left(\frac{\langle\,\log_{10}{[F_{\nu}(\lambda_i)]}\,\rangle-\log_{10}{[P_{\nu}(\lambda_i)]}}{\sigma(\langle\,\log_{10}{[F_{\nu}](\lambda_i)}\,\rangle)}\right)^2
\end{equation}

where $N$ is the number of datapoints being fit. Throughout this paper, the $\chi^2$ values mentioned are per datapoint.

This fitting process is repeated for a range of distances between $d_{\rm min}$ and $d_{\rm max}$, and a best-fit $d$, $\av$, and $\chi^2$ value can be found for each model SED.
The process is repeated for all the models in the grid. In this way, all the model SEDs in the grid are compared with the data, and the parameters of the model SEDs that fit a source well can be analyzed.

The process described above does not make any assumptions about the spatial extent of the source.
For example the apertures may be smaller than the extent of the source (see \S\ref{s:resolved} for an example).
Usually however, one measures the total flux of a source by using an aperture larger than the source.
With the knowledge that the source is not larger than a given aperture, we can cut down the number of model SEDs that fit its observed SED well.

For example, if a source in Taurus is unresolved at a given wavelength $\lambda_i$, then we should not be fitting models that would have been resolved at that particular wavelength.
More generally, if we know that the apparent extent of a source is always smaller than the apertures that the fluxes were measured in, then we want to eliminate all models whose apparent extent is larger than the aperture in at least one wavelength.
Note that we do not want to eliminate all models that are physically larger than the aperture, but only those that \emph{appear} larger than the aperture (the apparent size can be smaller than the physical size).
To use this information, we compute the wavelength-dependent outermost radius $R_{\lambda_i}(\Sigma=\Sigma_{0}/2)$ at which the surface brightness $\Sigma$ of the model is equal to half of the peak surface brightness $\Sigma_{0}$.
We can then reject all models that have $R_{\lambda_i}(\Sigma=\Sigma_{0}/2)~>~A(\lambda_i)$ in at least one wavelength $\lambda_i$.

The online version of the fitting tool is hosted on a dedicated Web
server\footnote{http://www.astro.wisc.edu/protostars}.
At the moment, it is possible to fit only a single source at a time, but we plan to upgrade this to allow users to upload data-files in order to fit multiple sources in late-2007.

\section{Proof-of-concept using known Taurus-Auriga sources}
\label{s:results}

\subsection{The data}
\label{s:data}
In this section we analyze whether by fitting model SEDs to observed SEDs we are able to correctly identify the evolutionary stage of YSOs, as well as determine the value of individual physical parameters.
To do this we have fit the SEDs of well studied and resolved YSOs in the Taurus-Auriga star-forming region.
Over the last few decades, Taurus has been the best studied star-forming region, and is in that respect the region for which the most data are available.
We have constructed a sample of 30 sources by selecting those from \citet{kh95} that have been spatially resolved as of mid-2006.
The latter requirement ensures that we have a good a-priori knowledge of the evolutionary stage of the objects from direct observations (e.g. young protostars with infalling envelopes, or disks).
FS~Tau and FS~Tau~B were not included due to their small angular separation, which makes it difficult to construct separate SEDs, and UZ~Tau E was not included because of insufficient SED coverage.
DG~Tau and DG~Tau~B were included, despite their small angular separation, as IRAC and MIPS data were available, allowing us to construct two separate well defined SEDs.

A few sources are known from direct imaging to be more complex than axisymmetric objects (e.g. IRAS~04016+2610; \citealt{padgett99} ; \citealt{wood01_4016}), but it is interesting to see what we can derive from fitting the SEDs of such sources.
Indeed, when looking at more distant star-forming regions, we have no a-priori knowledge of the complexity of the geometry of an unresolved source, and it is interesting to know what we can learn using 2-D models

We have compiled the SEDs for these 30 sources using optical (UBVRI), near- (JHKL) and mid-IR (MNQ) data from \citet{kh95}, JHK fluxes from the 2MASS point-source catalog \citep{skrutskie06}, IRAC data from \citet{hartmann05} and \citet{luhman06}, IRAS data from the IRAS point-source catalog and from \citet{weaver92} (the latter was used over the former when available), SHARC~II 350\micronsns, SCUBA 450\micronsns, and 850\microns sub-mm data from the compilation presented in \citet{andrews05}, and CSO 624 and 769\microns data from \citet{beckwith91}.

In addition, we have measured the IRAC fluxes for the sources not presented in \citet{hartmann05} or \citet{luhman06}, and re-measured the IRAC fluxes of RY Tau and DG Tau, which are saturated in all bands.
The IRAC data (PI Fazio - P00006 \& P00037) and the MIPS data (PI D. Padgett - P03584) were retrieved from the Spitzer Science Center Archive.
To measure the fluxes of sources saturated in IRAC, we used the publicly available iracworks code\footnote{http://spider.ipac.caltech.edu/staff/jarrett/irac} written by Tom Jarrett.
Finally, we have performed PSF photometry of MIPS~24 and 70\microns data, using a custom photometry code.
For sources that were mildly saturated in MIPS, we ignored the affected pixels
and ensured that the outer parts of the PSF were being correctly fit.

The list of sources along with the data are shown in Tables \ref{t:seds_1}, \ref{t:seds_2}, \ref{t:seds_3}, and \ref{t:seds_4} in Appendix~\ref{a:data}.

When fitting the SEDs, it is technically possible to use all the data available for each source, but in cases where several measurements were available at the same wavelength, we used the highest quality one.
For example, if MIPS data are available, we used it instead of IRAS, since IRAS has a poorer resolution, and could result in confusion.
Similarly, space-based IRAC data are usually preferred over ground based mid-IR data.
We applied the following rules:

\begin{itemize}
\item when 2MASS JHK data was available, it was used instead of previous JHK measurements.
\item when IRAC 3.6\,\microns data was available, the L-band flux was not used.
\item when IRAC 4.5 and/or 5.8\,\microns data were available, the M-band flux was not used.
\item when MIPS~24\,\microns data was available, the IRAS~25\,\microns flux was not used.
\item when MIPS~70\,\microns data was available, the IRAS~60\,\microns and 100\,\microns data were not used.
\item in general, flux measurements were always preferred over upper limits.
\end{itemize}

Additionally, we decided not to use the N and Q data when fitting the SEDs, due to large uncertainties in the fluxes, filter profiles, calibration, and zero-magnitude fluxes, but we quote these flux values in the tables for reference. 
A lower limit of 25\% was imposed on the flux uncertainties for optical, L \& M, IRAS, and sub-mm data, so that any flux measurement with a smaller uncertainty saw its uncertainty increased.
This was done to account for variability and uncertainties in the zero-magnitude fluxes in the optical, uncertainties in the filter profiles and zero magnitude fluxes in the L \& M bands, and uncertainties in the absolute flux calibration for IRAS and sub-mm wavelengths.
A lower limit of 40\% was imposed on the flux uncertainties for the CSO sub-mm data, as we did not have the transmission profiles for these observations: since the full width at half maximum (FWHM) of the filters used for the 769\,\microns observations is large (190\micronsns), differences in the filter profile would be noticeable.
For the transmission profiles of these two bands we used gaussians centered at 624 and 769\microns with FWHM 67 and 190\,\microns respectively, which we convolved with the atmospheric transmission profile used in \citet{dowell03}.
Additionally, a lower limit of 10\% was imposed on the remaining datapoints, to account mainly for uncertainties in the absolute calibration and photometry extraction.

\subsection{Comparison of derived physical properties to other methods}
\label{s:poc}

The only assumptions we made when fitting the observed SEDs with our model SEDs was that the sources were all situated in a distance range of $120\rightarrow160$AU (to rule out too luminous or too faint sources), that the foreground interstellar extinction was no more than $\av=20$ \citep{whittet01}, and we used the condition that none of the sources appeared larger than the apertures used to measure the fluxes.
The aperture radii assumed for the data are listed in Table \ref{t:seds_ap} in Appendix~\ref{a:data}.
%We could be more strict in the latter case, since many of the sources are no larger than $\sim10''$ at the very most (IRAS~04368+2557 however does extend out to $\sim100''$).
%However, when analyzing more distant star-forming regions, we may not be able to constrain as easily the physical size of the objects.
%For example, an upper limit on the angular size of $\sim3''$ at $2.5$kpc corresponds to $\sim50''$ at the distance of the Taurus-Auriga complex.
%Therefore, by not placing any strict constraint on the angular extent of the sources, we are adopting the same conditions that would be present in observations of more distant star-forming regions.

\begin{figure}
\epsscale{0.80}
\plotone{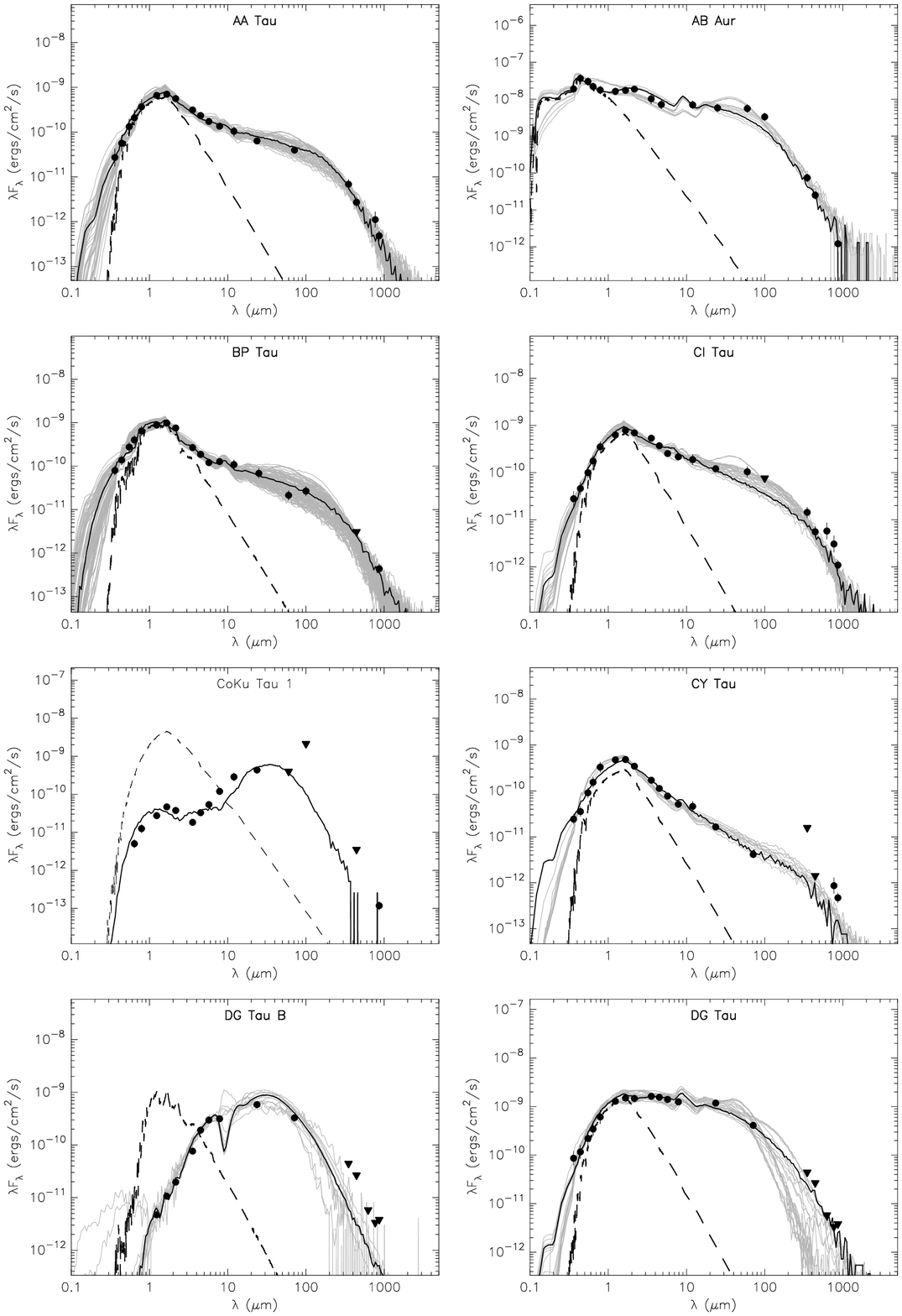}
\caption{\small SEDs for the 30 Taurus-Auriga sources analyzed in this paper.
Filled circles are the flux values listed in Tables \ref{t:seds_1},
\ref{t:seds_2}, \ref{t:seds_3} and \ref{t:seds_4}. Triangles are upper limits.
Error bars are shown if larger than the data points. The solid black line
indicates the best fitting model, and the gray lines show all models that also
fit the data well (defined by $\chi^2-\chi^2_{\rm best}<3$, where $\chi^2$ is
the value per datapoint). The dashed line shows the SED of the stellar
photosphere in the best fitting model.\label{f:seds}}
\end{figure}

In Figure \ref{f:seds} we show the SEDs for all the sources along with the best fitting model for each source, and all the models with $\chi^2-\chi_{\rm best}^2<3$ (where $\chi^2_{\rm min}$ is the $\chi^2$ per datapoint of the best fitting model for each source). Although this cutoff is arbitrary, it provides a range of acceptable fits to the eye.
A purely statistical argument would of course be more desirable, but the sampling of our models in 14 dimensional space is too sparse to resolve well enough the shape of the minima in the chisquared `surface' to derive formal confidence intervals. 
Using a `chi-by-eye' approach does carry the risk of overestimating the uncertainties on the various parameters, but we argue that this is better than under-estimating the uncertainties, which would lead to over-interpretation.
Furthermore, intrinsic variability of the sources, along with the fact that no young stellar object will be as perfectly symmetrical and well behaved as our models, means that any `formal' confidence interval would likely be too strict.

All sources except DM Tau and GM Aur have their SED reasonably well fit by our model SEDs. These sources are known to show a near-IR deficit in their SEDs \citep{rice03,calvet05}, suggesting that the inner region of their disks has been cleared of dust. Although we do include a substantial number of models with inner holes in our grid, our models systematically over-estimate the observed mid-IR fluxes. This may be because we evacuate the disks completely below a given radius, rather than having small but non-zero amounts of dust inside the `hole'.  We plan to address this in a future grid of models. This is an example where fitting model SEDs to data provides us with feedback to improve our models.

The SED for CoKu~Tau/1 is fit by only one model within the goodness-of-fit
constraints. The model that does fit well is that of an edge-on YSO with a
remnant accreting envelope (see Table \ref{t:mdot}). Based on the SED fit alone,
an edge-on inclination seems reasonable, as the SED is double-peaked, a typical
signature of an edge-on disk. \citet{stark06} modeled spatially resolved near-IR
images of CoKu~Tau/1, and found that it does indeed requires a low-mass
envelope, suggesting that CoKu~Tau/1 is more evolved than a typical Class I
source. However, they find that the images are best-fit using a 64$^{\circ}$
inclination rather than edge-on. This seems more plausible as the central source
is visible in these images (the central source would not be visible in an
edge-on disk). The central source is a binary system with separation 0.24''
($\sim$33\,AU at 140\,pc), which will likely have evacuated some material from
the inner disk. At this point, we can only speculate about the reasons for the
discrepant results between the SED and image fitting:  Perhaps one of the
sources in the binary is surrounded by a small close to edge-on disk inside the
larger circumbinary disk. This may explain why one of the sources in the binary
is redder in NICMOS images than the other source. We plan to further model the
multi-wavelength images of CoKu~Tau/1 to test this hypothesis.

In general, we note that the main effect a young binary system will have on its
circumstellar material will be to clear out, at least partially, the innermost
regions of the disk or envelope of gas and dust.  In fact, this is the primary
reason that large inner holes are included in our models. The main issue will
likely be the determination of the properties of the central sources. In cases
where the binary separation is large enough (e.g. CoKu~Tau/1 or GG~Tau), the
geometry will become more complex, with a circumbinary disk, and possibly small
individual disks around each source. In these cases, the disk parameters will
also be wrong, as the geometry is not being modeled correctly. In the future we
plan to explore in more detail the effects of a binary system on the
determination of the physical parameters of YSOs.

In the remainder of this section, we compare a selection of physical properties
of the sources derived from fitting model SEDs to the sources with values quoted
in the literature:

\paragraph{Evolutionary Stage and Viewing Angle} The first result that we can
compare with known values is whether the correct evolutionary stage is
identified, i.e. whether young protostars with infalling envelopes or disks are
correctly identified as such (see Table \ref{t:mdot}).
To do this, we look at the range of envelope accretion rates that provide a good
fit to each source, and specifically the lower value of the range of
good-fitting models.
If this lower value is greater than zero, then this tells us that the source \emph{cannot} be explained by a disk-only source within the goodness-of-fit constraints applied.
Conversely, if the lower value is zero, then this tells us that the source can be explained by a disk with no infalling envelope.
In most cases where the lower value is zero, the upper value is non-zero, which means that it is impossible to rule out that there may be very optically thin envelopes surrounding the disks.
We also list the ranges of inclinations to the line-of-sight
that provide a good fit as there is in some cases a degeneracy between
evolutionary stage and inclination.
We can make several remarks about the results:

\begin{itemize}
\item All known disk sources (marked as `Disk' in Table \ref{t:mdot}) except DM Tau are well fit by models with no infalling envelope.
Furthermore, the maximum value of the accretion rate providing a good fit for these sources does not exceed $10^{-6}\msol$\,yr$^{-1}$.
This is expected, as we showed in Paper I that envelopes with accretion rates lower than $\sim10^{-6}\msol$\,yr$^{-1}$ do not contribute significantly to the SED.
% REMOVE THIS

\item Most sources that are known to have infalling envelopes (marked as `Embedded' in Table \ref{t:mdot}), cannot be fit by disk-only models.

\item The exceptions to the above point are DG Tau B, IRAS~04016+2610 and
IRAS~04248+2612, which are also known to have infalling envelopes, but can be
fit by disk-only models as well as models with infalling envelopes. However, for
all three sources, the disk-only models that provide good fits are all viewed
edge-on (the lower limit on the range of inclinations for
disk-only models, i.e. models for which $\mdote=0$, which fit these sources
well is larger than 80$^{\circ}$). This is a good
demonstration that even using an SED with data ranging from optical or near-IR
to sub-mm is not always enough to unambiguously distinguish between various
geometries once viewing angle is taken into account.
%\item CoKu~Tau/1, which is a disk with a remnant envelope \citep{stark06}, is well fit only by models with no envelope. This is plausible, as a remnant envelope will not contribute significantly to the SED.
\end{itemize}

In practice, one could eliminate model fits based on statistical arguments.
For example, in a star-forming cluster, observing a source with an edge-on disk has a low probability both because one is less likely to observe a disk with an edge-on inclination, and because such a source would be much fainter.
As discussed further in this section, the edge-on disk models that fit
the observed SEDs well require the luminosity of the central star to be larger
than 100\,$\lsol$, which is unrealistic, as we would in this case expect to
observe many more non edge-on sources with such luminosities.

We note that in most cases, the inclination of the line of sight
is not well determined. The only constraints we find are that none of the
disk-only sources are seen exactly edge-on (87$^{\circ}$ in Table 1), and that
CoKu~Tau/1 cannot be fit by any non edge-on models as discussed previously.

\begin{figure}
\includegraphics[angle=270,width=6in]{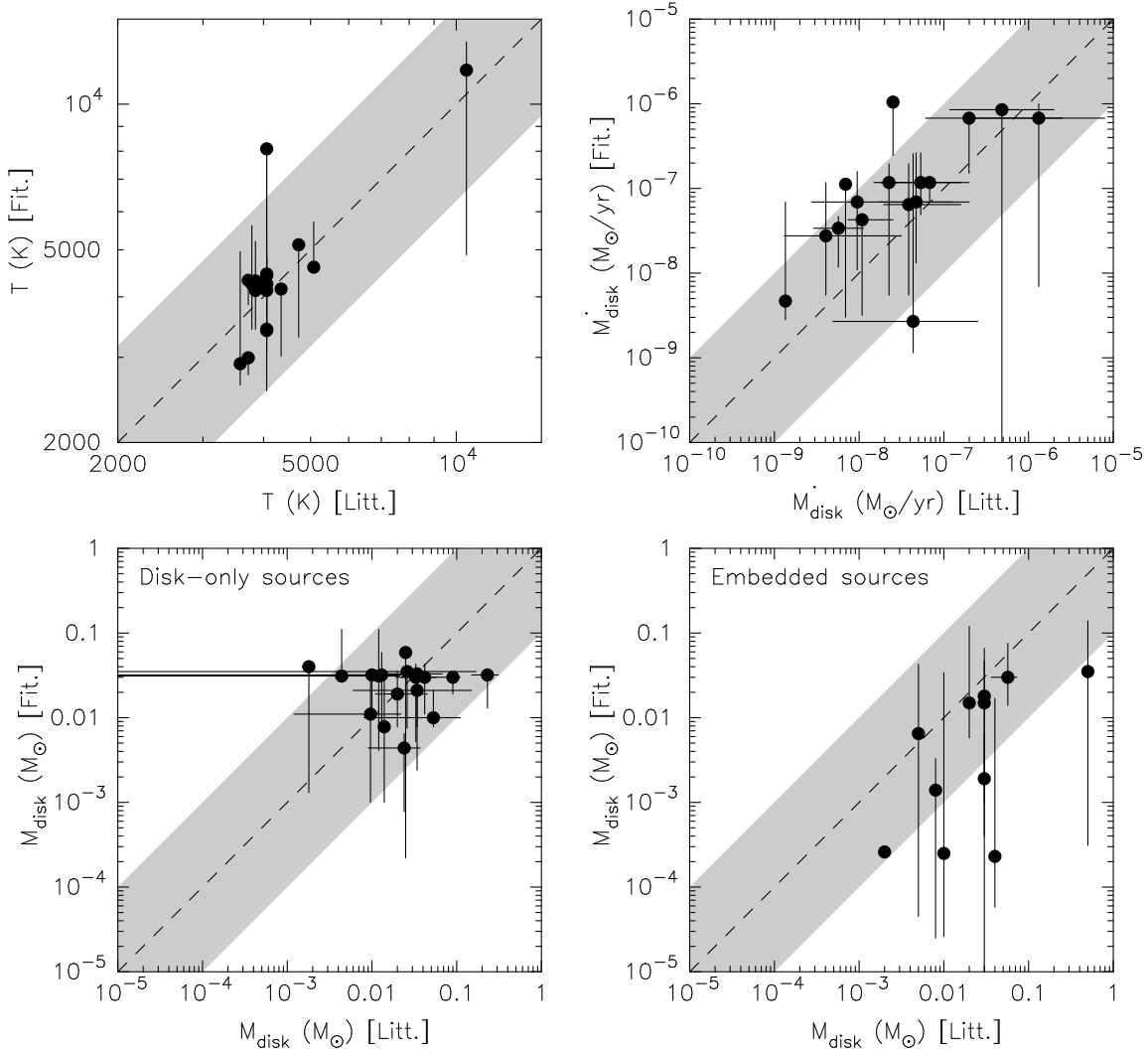}
\caption{Comparison of the values of stellar temperature (top left), disk accretion rate (top right), and disk mass (bottom left and right) found from SED fitting with values quoted in the literature and derived from different methods. The dashed line shows where the values would be equal, and the gray area shows where the agreement is better than $\pm0.2$ orders of magnitude for the temperature, and $\pm1$ order of magnitude for the disk mass and accretion rate. Error bars in the y-direction indicate the range of values of the models shown in Figure \ref{f:seds}. Error bars in the x-direction show the range between the lowest value and the highest value quoted in the literature. For the stellar temperature, only one reference \citep{kh95} was used, and therefore no uncertainties are shown on the x-axis. The likely uncertainty quoted by the authors is $\pm 1$ spectral class, corresponding roughly to $\pm100$K.\label{f:params}}
\end{figure}

\paragraph{Stellar Temperature} In Table \ref{t:temp} and in Figure \ref{f:params} we compare the temperatures derived from SED fitting to those corresponding to the spectral types listed in \citet{kh95}.
The latter were determined from spectroscopic observations, and therefore provide an independent and more accurate measure of the temperature.
We find that for all sources but one, the best-fit value is close to the known value.
The exception to this is UY Aur, but we note that the \emph{range} of temperatures for this source ($3540\rightarrow8090$\,K) is consistent with the literature value ($4060$\,K).

AB Aur is the only source in our sample that is known to have a high temperature ($\sim10,500$\,K), and we note that it is also the source for which the best-fit model has the highest temperature (11,767\,K).
%Additionally, we cannot fit the SED of AB Aur with any models with temperatures below 10,726\,K.
Generally, we find that we have correctly identified the temperatures of the sources to better than $\pm0.2$ orders of magnitude.

\paragraph{Disk Mass} We compare the disk masses derived from SED fitting to values listed in \citet{dutrey96}, \citet{kitamura02}, and \citet{andrews05}.
These values were determined from sub-mm and mm observations.
%, and are therefore determined using a method more similar to our own.
In Table \ref{t:mdisk} we list the various literature values, including an average value, a minimum value defined to be the smallest value of $\mdisk-\sigma(\mdisk)$ quoted, and a maximum value defined to be the largest value of $\mdisk+\sigma(\mdisk)$ quoted.
We compare this range to the range found from SED fitting (this is also shown graphically in Figure \ref{f:params}).
Our values assume a gas-to-dust ratio of 100.
As can be seen from Figure \ref{f:params}, for the very young sources (marked
`Embedded' in Table \ref{t:mdot}) we do not constrain the disk mass well.
For example IRAS~04361+2547 could, on the basis of the SED, have a disk mass between $2.5\times10^{-5}$\,$\msol$ and $3.2\times10^{-2}$\,$\msol$.
This is because, in the early stages of evolution, when the disk is deeply embedded inside the infalling envelope, the relative contributions of the disk and envelope to the SED are difficult to disentangle.
%This is expected, as in early stages of evolution the disk does not contribute to a large fraction of the emission from the source, since it is deeply embedded inside the infalling envelope (\citealt{whitney03p1}, Paper I).
For disk-only sources (marked `Disk' in Table \ref{t:mdot}), we find that our values generally agree with the literature values to within less than an order of magnitude.
This is fairly accurate considering that the disk masses in our model grid range over nine orders of magnitude, and that literature values often rely on simple models (e.g. power laws), and assume that the sub-mm flux arises from an optically thin isothermal region of disks.
Furthermore, the choice of a dust opacity law will also affect the disk mass obtained.

\paragraph{Disk Accretion Rates} In Table \ref{t:mdotdisk} and in Figure \ref{f:params} we compare the disk accretion rates derived from SED fitting to values listed in \citet{valenti93}, \citet{hartigan95}, \citet{hartmann98}, and \citet{mohanty05}.
Most of these literature values were derived from UV and optical spectroscopy, and therefore represent an independent and more accurate estimate of the disk accretion rate than SED fitting.
We find that the agreement between values derived from SED fitting and literature values is reasonable over the three orders of magnitude spanned.

The values obtained from SED fitting appear to be slightly larger than those taken from the literature.
This could be due to a small inconsistency in our models where the accretion luminosity inside the dust destruction radius but outside the magnetic truncation radius is, for simplicity, being emitted as a stellar photon, with a resulting stellar spectrum.
Thus, to match a given near-IR excess, a slightly higher accretion rate is required.
For example, RY~Tau is the source for which our estimate deviates the most from literature values; \citet{akeson05} found that for this source, the accretion luminosity from the gas disk inside the dust destruction radius contributed significantly to the total disk accretion luminosity.
Future versions of our grid of models will include disk emission inside the dust destruction radius.

\paragraph{Disk/Envelope Inner Radius} All the sources in the sample can be fit by disks and envelopes with no inner holes ($\rmind=\rsub$ where $\rsub$ is the dust sublimation radius defined in Paper I), with the exception of DM Tau, GM Aur, IRAS 04302+2247, and IRAS04325+2402 (see Table \ref{t:addpar}).
As described in Paper I, the inner radius is the same for the disk and the envelope.
For most sources, the upper limit to the disk inner radius is larger than $\rsub$, and it is generally not possible to rule out the existence of holes in any of the sources.
However, in the case of DM Tau, GM Aur, IRAS 04302+2247, and IRAS04325+2402, the lower limit being larger than $\rsub$ \emph{does} rule out disks or envelopes with \emph{no} inner holes.
GM Aur has previously been found to require an inner hole of order $\sim4$AU in order to explain its SED \citep{rice03}.
More recently, {\it Spitzer} IRS spectroscopy has confirmed that both GM Aur and DM Tau show a near-IR deficit, suggesting inner hole sizes of $24$\,AU and $3$\,AU respectively \citep{calvet05}.
We fit the SED of GM Aur with models having an inner hole size between $\sim1$ and $24$ AU, and DM Tau with models having an inner hole size between $\sim3.4$ and $20$\,AU: these values are consistent with the values found by \citeauthor{calvet05}.

The above comparisons show that in most cases, the parameters of the best fit model agree well with previously published values.
Even in cases where the best-fit value does not agree with the published values, the \emph{range} of values providing a good fit is consistent with previously published values.

In addition to the parameters presented above, we also present the ranges of values providing a good fit for the stellar mass, bolometric luminosity, disk outer radius, and disk scaleheight $h$ at 100\,AU.
The values are presented in Tables \ref{t:addpar} and \ref{t:addpar2}.
The disk outer radii are generally fairly uncertain, and range from tens to hundreds of AU.
The stellar masses are likely to be constrained because we determine temperatures well, and our model grid uses evolutionary tracks, relating the stellar temperatures to the stellar masses.
We do not have literature values to compare the derived scale heights to. However we do expect the derived values to be consistent with the true values: as shown in Paper I, the disk flaring power does have an effect on the SED, meaning that we should be somewhat sensitive to the disk scaleheight. The values are poorly constrained for embedded sources, where the disk does not contribute significantly to the SED.
Finally, the main uncertainty in the bolometric luminosity is due to uncertainties in the inclination.
For example the sources for which the luminosity is very poorly constrained (from a few $\lsol$ to over 100\,$\lsol$), such as DG Tau B or IRAS 04016+2610, are those for which we cannot distinguish between an edge-on disk or a young protostar with an infalling envelope.
These two possibilities would result in two very different bolometric luminosities.
If the inclination can be constrained, then we can potentially determine the true luminosity of the source very accurately from SED fitting.

\subsection{Constraining parameters: uniqueness of fits, degeneracies, and wavelength coverage}

\label{s:degen}

As mentioned in Section~\ref{s:intro}, when fitting model SEDs to multi-wavelength observations of a source, it is important to have a grasp not only of which set of parameters provides a good fit, but also whether it is the \emph{only} set of parameters that does so.
For example, a source may be well fit by a model with a 1\,$\msol$ central source and a $300$\,AU disk with a mass of $\mdisk=0.01\,\msol$ disk, but it may be that the radius of the disk does not actually have an effect on the SED at the wavelengths of the available data, and that any disk radius would provide a suitable fit.
Furthermore, two completely different sets of parameters may fit data equally well (e.g. the sources from Section \ref{s:poc} that can be fit by SED models of young protostellar objects or older edge-on disks).

The reader may wonder how we can claim to determine 14 parameters from fitting SEDs to data that may or may not span a large wavelength range.
The answer is that we do not claim to \emph{determine} any of the parameter values, but we aim to find \emph{how well constrained} each parameter is.
For example, from fitting IRAC data alone, we may find that none of the parameters are well constrained and that adding an extra datapoint only allows us to determine for example an upper limit on one of the parameters.

In this section, we show how the YSO model SED fitting tool can be used to analyze how well various parameters are constrained when using different combinations of datapoints.
In addition, we show examples of degeneracies that can arise in parameter space.
This is by no means an exhaustive study, and we encourage readers to use the online fitting tool to explore in further detail how well various parameters are constrained for a specific set of data.

\begin{figure}
\epsscale{0.90}
\plotone{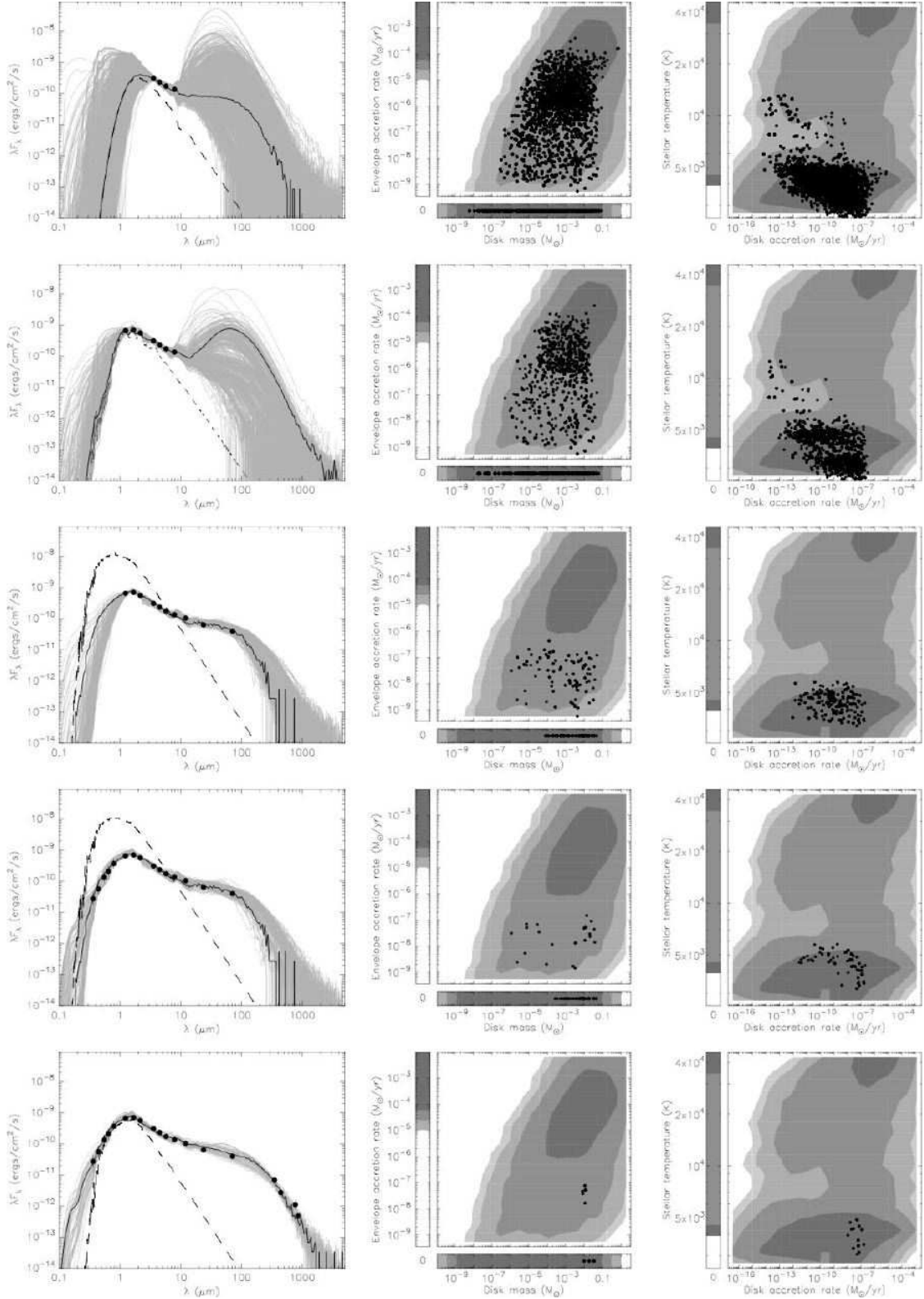}
\caption{\scriptsize Left: model SED fits to the observed SED of AA Tau using
from top to bottom: IRAC points only, JHK + IRAC points, JHK + IRAC +
MIPS~24\,$\mu$m + MIPS~70\,\micronsns, UBVRI + JHK + IRAC + MIPS~24\,$\mu$m +
MIPS~70\,\micronsns, and UBVRI + JHK + IRAC + MIPS~24\,$\mu$m + MIPS~70\,$\mu$m
+ sub-mm points (when more than 200 model SEDs fit the data well, we show the
200 best fit SEDs, and one in ten SEDs beyond this). Center and right: a
selection of parameters for the model fits. Filled circles are the good fits
shown in the SED plots. The grayscale shows the parameter space of the grid of
models (smoothed). Each shade of gray shows an increase in a factor of 10 in the
density of models.
\label{f:source1}}
\end{figure}

The first example that we show is AA Tau, which is a known T-Tauri source with optical variability.
Figure \ref{f:source1} shows the model SED fits to the data, and a selection of parameters for these fits (disk mass and accretion rate, envelope accretion rate, and stellar temperature).
For each combination of datapoints we show all models with $\chi^2-\chi^2_{\rm min}<3$ as in Section \ref{s:poc}.
As before, we assume a distance range of $120\rightarrow160$pc.
We first fit the SED using only IRAC points, and we find that 11,674 model SEDs satisfy the goodness of fit criterion ($\sim5$\% of all model SEDs!).
The disk mass, disk accretion rate and envelope accretion rate are not at all constrained apart from very high values of the temperature being ruled out (this is because models with very high temperatures would be too luminous to explain the IRAC fluxes).
This suggests that using IRAC fluxes alone does not necessarily provide a good estimate of the evolutionary stage of an object.
Adding JHK data reduces the number of good fits to 2,826 model SEDs, but does not provide any significant improvement in the determination of the four parameters that we show.
Adding IRAS~12\,\micronsns, MIPS~24\,\microns and MIPS~70\,\microns does have an important effect which is to provide a much more strict upper limit on the envelope accretion rate ($\sim10^{-7}\,\msol$\,yr$^{-1}$), and also provides a lower limit on the disk mass of $10^{-6}\msol$.
Furthermore, the disk accretion rate and the central source temperature are better determined.
This echoes our findings presented in Paper I that data beyond $\sim20$\,\microns is very helpful in constraining the evolutionary stage of a source.
Adding the optical data rules out models that have both low central source temperatures and low disk accretion rates, leaving a degeneracy between the optical data being explained by a higher temperature or a higher disk accretion rate.
Finally, adding the sub-mm data has the predictable effect of providing a strong constraint on the disk mass, placing it at $\sim10^{-2}\msol$.
The degeneracy between central source temperature and disk accretion rate is resolved, and the disk accretion rate is estimated at $\sim10^{-8}-10^{-7}\msol$/yr.

\begin{figure}
\epsscale{0.90}
\plotone{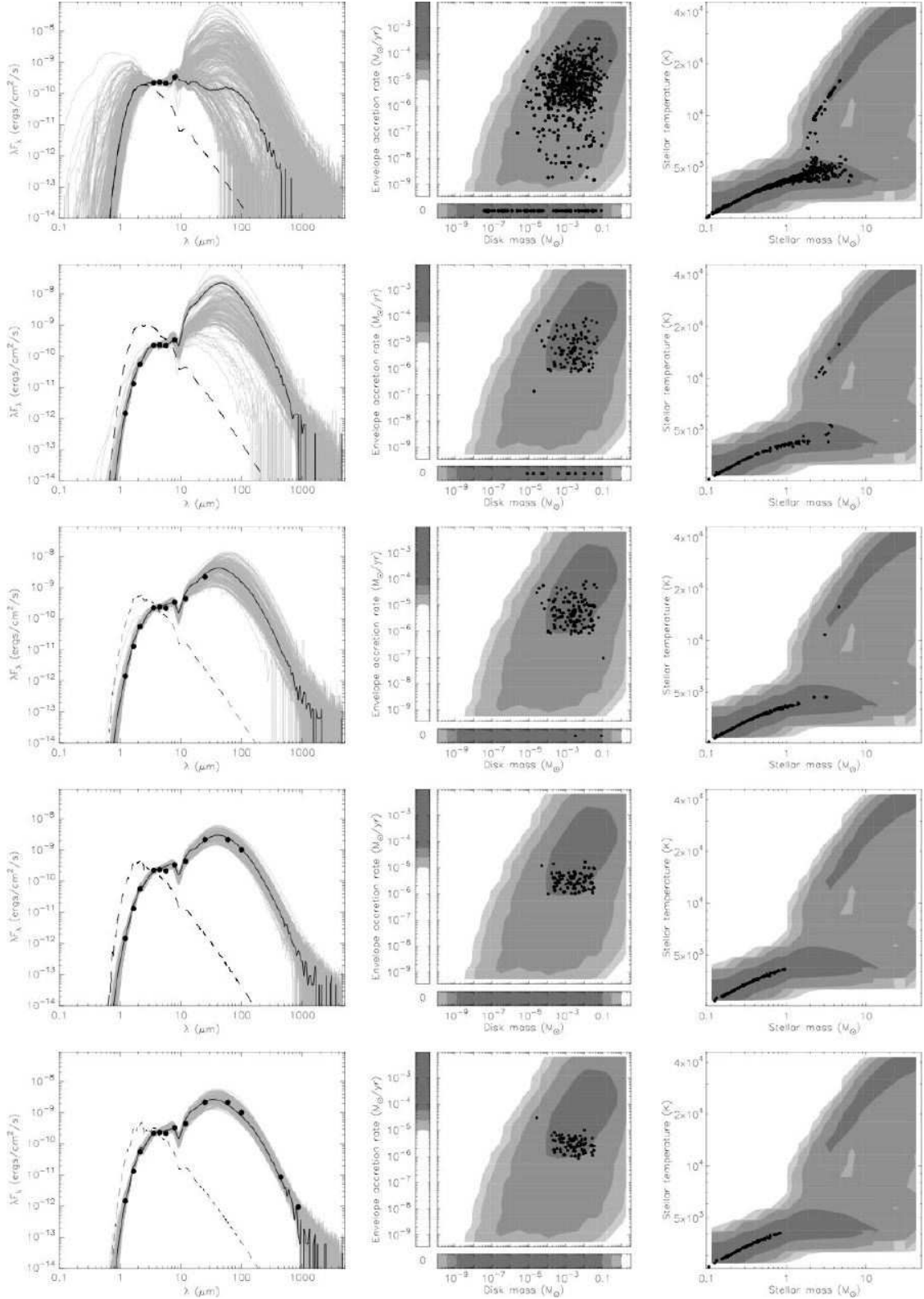}
\caption{\scriptsize Left: model SED fits to the observed SED of IRAS~04361+2547
using from top to bottom: IRAC points only, JHK + IRAC points, JHK + IRAC +
IRAS~12 \& 25\,\micronsns, JHK + IRAC + All IRAS bands, and JHK + IRAC + All
IRAS bands + sub-mm points (when more than 200 model SEDs fit the data well, we
show the 200 best fit SEDs, and one in ten SEDs beyond this). Center and right:
a selection of parameters for the model fits. Filled circles are the good fits
shown in the SED plots. The grayscale shows the parameter space of the grid of
models as for Figure \ref{f:source1}. The alignment of all the fits along a
curve in the $\tstar$ vs $\mstar$ plots (right) is due to the sampling of the
models using evolutionary tracks, i.e. for a given $\mstar$ and $\agestar$ we
find $\tstar$ using evolutionary tracks (see Paper I).\label{f:source2}}
\end{figure}

Our second example is IRAS~04361+2547, which is a known protostar surrounded by an infalling envelope and with molecular outflows, as shown by spatially resolved observations \citep[e.g.][]{terebey90,tamura91}.
Figure \ref{f:source2} shows the model SED fits to the data, and a selection of parameters for these fits (disk mass, envelope accretion rate, and stellar mass and temperature).
As for AA Tau we show all models with $\chi^2-\chi^2_{\rm min}<3$.
We first fit the SED using only IRAC points, and we find that 1,959 model SEDs satisfy the goodness of fit criterion.
As before, the disk mass and envelope accretion rate are not at all constrained.
Stellar mass and temperature are also very poorly constrained. 
Adding JHK data reduces the number of good fits to 183, and does provide an improvement in the determination of the envelope accretion rate.
However, we note that there is a bimodal distribution of models in ($\mdote,\mdisk$), with one peak centered at fairly high accretion rates ($\sim10^{-5}\msol$/yr), and one centered at zero accretion rates (which are mostly edge-on disk-only models).
A bimodal distribution is also seen in ($\tstar,\mstar$).
The models with high temperatures correspond to the edge-on disk-only models, which require on average a larger central source luminosity to match the observed fluxes, while the models with lower temperatures correspond to the young models with large envelope infall rates.
This is a good example of a young protostar - edge-on disk degeneracy.
The addition of IRAS~12\,\microns and IRAS~25\,\microns reduces the number of edge-on disk models that fit well, but does not rule them out.
Adding IRAS~60\,\microns and IRAS~100\,\microns rules out all edge-on disk models, and in the process rules out all models with $\mstar>1\msol$.
Finally, adding the sub-mm data further constrains the parameters further, albeit not much.
By this stage, the SED fit is fairly well constrained.
The envelope accretion rate is constrained to within an order of magnitude, the disk mass is determined to within two orders of magnitude, the temperature is determined to be no more than 4,000\,K, and the stellar mass is determined to be no more than 1\,$\msol$.

\subsection{Analysis of resolved sources}

\label{s:resolved}

As mentioned in Section \ref{s:tech}, when analyzing the SED of a resolved object, it is possible to specify apertures smaller than the apparent extent of the source.
In this section, we demonstrate this by modeling IRAS~04368+2557 using our model SED fitter and using solely IRAC data.

\begin{figure}
\epsscale{0.80}
\plotone{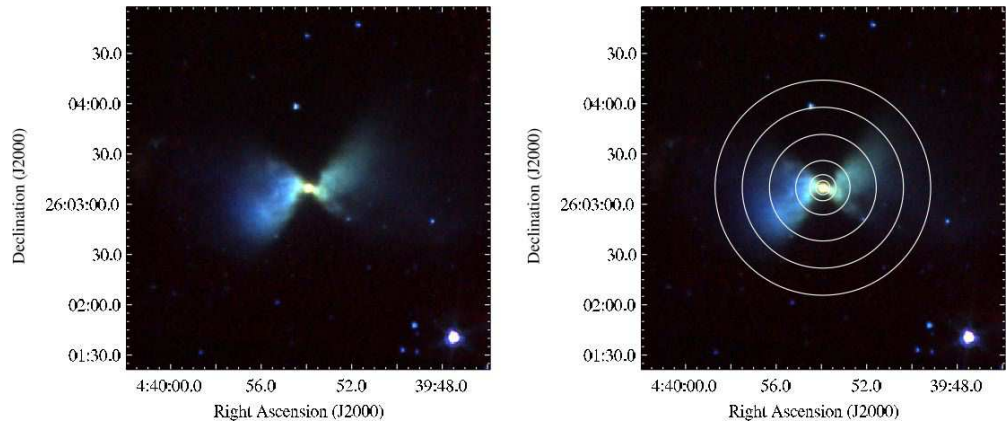}
\caption{Left: Three color image of IRAS~04368+2557 using IRAC 3.6\microns (blue - 0 to 3MJy.sr$^{-1}$), IRAC 4.5\microns (green - 0 to 6MJy.sr$^{-1}$), and IRAC 7.8\microns (red - 0 to 12MJy.sr$^{-1}$). Right: The same image with the six photometry apertures overplotted (4'', 8'', 16'', 32'', 48'', and 64''). \label{f:04368d}}
\end{figure}

In order to model IRAS~04368+2557, we found the total flux in the four bands in six different apertures: 4'', 8'', 16'', 32'', 48'', and 64''.
A composite color image of the source is shown in Figure \ref{f:04368d}, along with the position of the six apertures.

We imposed a distance constraint of 120 to 160pc to the object, as well as an Av to the object of 0 to 20.
We did \emph{not} impose any conditions such as an inclination of the object to the line of sight, nor the cavity size, as we wish to find whether the fitting tool can determine such parameters solely from the aperture information.

\begin{figure}
\epsscale{1.00}
\plotone{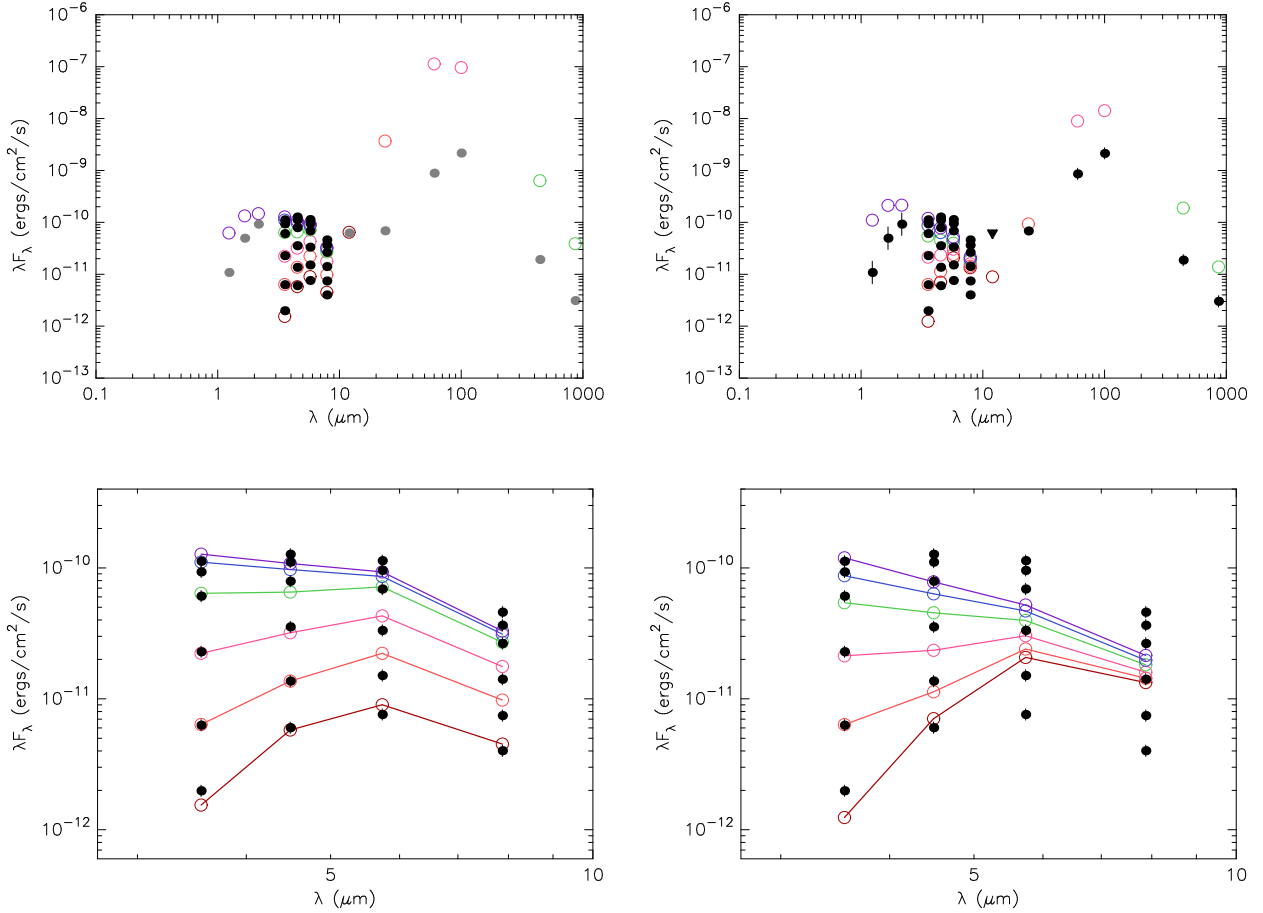}
\caption{Top left: best-fit model to the IRAC data in the six apertures for
IRAS~04368+2557 (only the IRAC data is used). The IRAC datapoints are shown as
filled black circles. The other datapoints are shown as gray circles to indicate
that these were not used in the fitting. The apertures assumed for these points
are those listed in Table \ref{t:seds_ap} as before, i.e. 100'' for JHK, 10''
for \mipsns, 60'' for IRAS 12 \& 25\,$\mu$m, 120'' for IRAS 60 \& 100\,$\mu$m,
and 30'' for the sub-mm data.
For the IRAC datapoints, the fluxes in six apertures are shown (4'', 8'',
16'', 32'', 48'', and 64''), with the faintest fluxes corresponding to the
smallest apertures, and the brightest fluxes corresponding to the largest
apertures.
The model SED in
the different apertures is shown as open circles. The different colors
correspond to the different apertures. Top right: best-fit model using the
near-IR, far-IR, and sub-mm data in
addition to the IRAC data in the six apertures. The bottom left and right panels
show the same SED fits in the IRAC range. \label{f:04368sed}}
\end{figure}

\begin{figure}
\epsscale{0.30}
\plotone{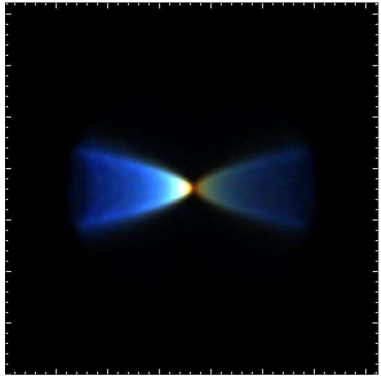}
\caption{Image of the best-fit model to the IRAC data only for IRAS\,04368+2557,
shown in the same bands (IRAC 3.6, 4.5, and 7.8\,$\mu$m) and with the same
scaling as the image in Figure \ref{f:04368d}. \label{f:04368m}}
\end{figure}

The left plots in Figure \ref{f:04368sed} shows the fluxes for the six apertures along with the best fitting aperture-dependent model SED.
Also shown are the near-IR, far-IR, and sub-mm datapoints, but these are initially not used in the fitting process.
Using the parameters of the best fit, we produced a model image, shown in Figure \ref{f:04368m}, by rerunning this model with a higher signal-to-noise.
The model SED fitter correctly finds an SED that matches the data adequately in all apertures, determines the type of object (infalling envelope), the inclination to the line of sight (edge-on), and the colors of the object. The cavity angle is not a perfect match, but this is not surprising since we only provided the fitting tool with circularly averaged fluxes, containing no information about any possible asymmetries.
 
We note however that all the models that fit the resolved IRAC data well systematically overestimate fluxes at far-IR wavelengths (as seen in the top left plot of Figure \ref{f:04368sed}).
The plots on the right of Figure \ref{f:04368sed} show that fitting the data at other wavelengths simultaneously deteriorates the fit at IRAC wavelengths.
Furthermore, our models do not reproduce the peak in luminosity at the center of the object (as seen in Figure \ref{f:04368m}).
It has recently been found \citep{loinard02} that the central source of IRAS~04368+2557 is a binary star with a projected separation of 25\,AU, meaning that we expect a large cavity at the center of the infalling envelope.
As mentioned previously, although our model grid does include models with inner holes, we evacuate the holes completely, rather than leaving low levels of dust.
A small but non-zero amount of dust inside an inner hole may be enough to provide the IRAC fluxes observed, while decreasing the amount of far-IR flux.
The parameters of the best-fitting models for the IRAC data
only, and for the full SED, are listed in Table \ref{t:04368tab}.

As well as demonstrating SED fitting of resolved sources, this is an illustration of how detailed modeling of an individual source can be used to improve our models for a future grid. In this future grid we also plan to produce images at a number of different wavelengths for all the models in the grid, allowing us to fit multi-wavelength images of resolved YSOs without first averaging the flux in circular apertures.

\section{Summary and Conclusions}

We have developed a method to fit observed YSO SEDs using a pre-computed grid of models.
Although we vary the values of 14 physical parameters in our models (c.f. Paper I), we do not claim to be able to determine all of these parameters from fitting observed SEDs.
Instead, we are interested in determining which parameters can be contrained, if at all, and if so, what range of values provide acceptable fits.
We have compiled data from UBVRI to sub-mm wavelengths for 30 YSOs in the Taurus-Auriga star-forming region.
The evolutionary stages, stellar temperatures, disk masses, and disk accretion rates derived from fitting the model SEDs to the data are in good agreement with independently determined values (for example from spectroscopy).
In cases where the best fitting model is not in agreement with the literature value, the \emph{range} of parameter values of the model SEDs that provide a good fit is in general still consistent with this value.
We have also demonstrated how adding fluxes at various wavelengths helps constrain different parameters.
For example we find, as in Paper I, that data in the range 20-100\microns in addition to shorter wavelength data is very useful in constraining parameters such as the envelope accretion rate, and thus the evolutionary stage of a source.
An online fitting tool making use of this method is available to the
community\footnote{http://www.astro.wisc.edu/protostars}.
In future, we plan to fit spectra (e.g. {\it Spitzer} IRS data) and polarization measurements simultaneously with the broadband SED, in order to place further constraints on the geometry and the chemistry of the circumstellar environment of YSOs.

\acknowledgements

We wish to thank Ed Churchwell, Martin Cohen, Keith Horne, and Katharine Johnston for useful discussions, Mike Wolff for models of the dust extinction law, and Stephan Jansen for help with setting up the web server for the online model grid and SED fitting tool.
We also wish to thank the anonymous referee for suggestions which helped improve this paper.
Partial support for this work was provided by a Scottish Universities Physics Alliance Studentship (TR), and a PPARC advanced fellowship (KW).
Additional support, as part of the Spitzer Space Telescope Theoretical Research Program (BW, TR), Legacy Science Program (BW, TR, RI), and Fellowship Program (RI), was provided by NASA through contracts issued by the Jet Propulsion Laboratory, California Institute of Technology under a contract with NASA.
This research has made use of the SIMBAD database, operated at CDS, Strasbourg, France; is based in part on observations made with the Spitzer Space Telescope, which is operated by the Jet Propulsion Laboratory, California Institute of Technology under a contract with NASA; and has made use of data products from the Two Micron All Sky Survey, which is a joint project of the University of Massachusetts and the Infrared Processing and Analysis Center/California Institute of Technology, funded by NASA and the National Science Foundation

%%%%%%%%%%%%%%%%% References %%%%%%%%%%%%%%%%%

%\bibliographystyle{apj}
%\bibliography{apj-jour,references} 
\bibliography{}
  
%%%%%%%%%%%%%%%%% Figures %%%%%%%%%%%%%%%%%
\clearpage
\newpage

\clearpage
{\plotone{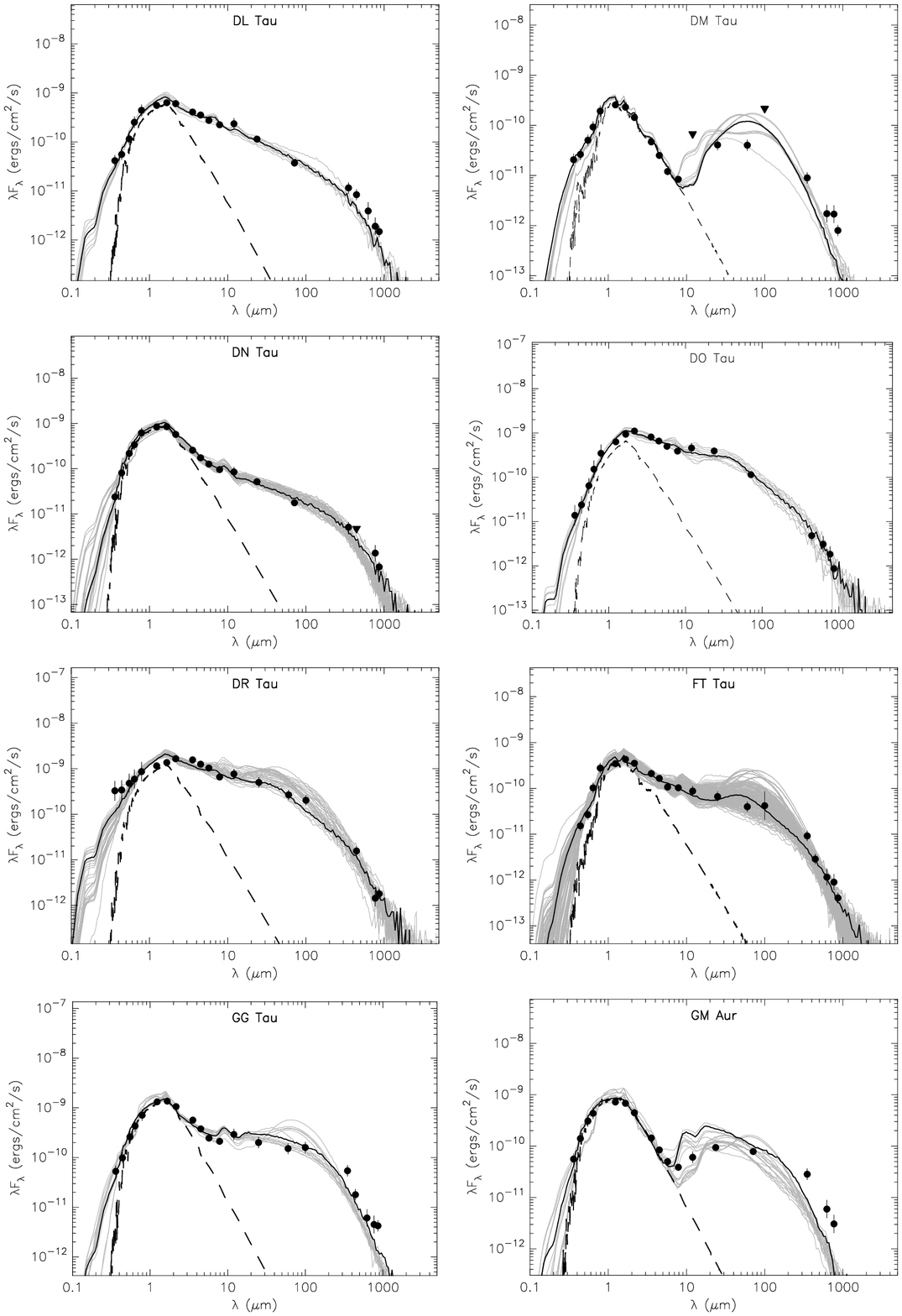}}\\[5mm]
\centerline{Fig. 1. --- Continued.}
\clearpage
{\plotone{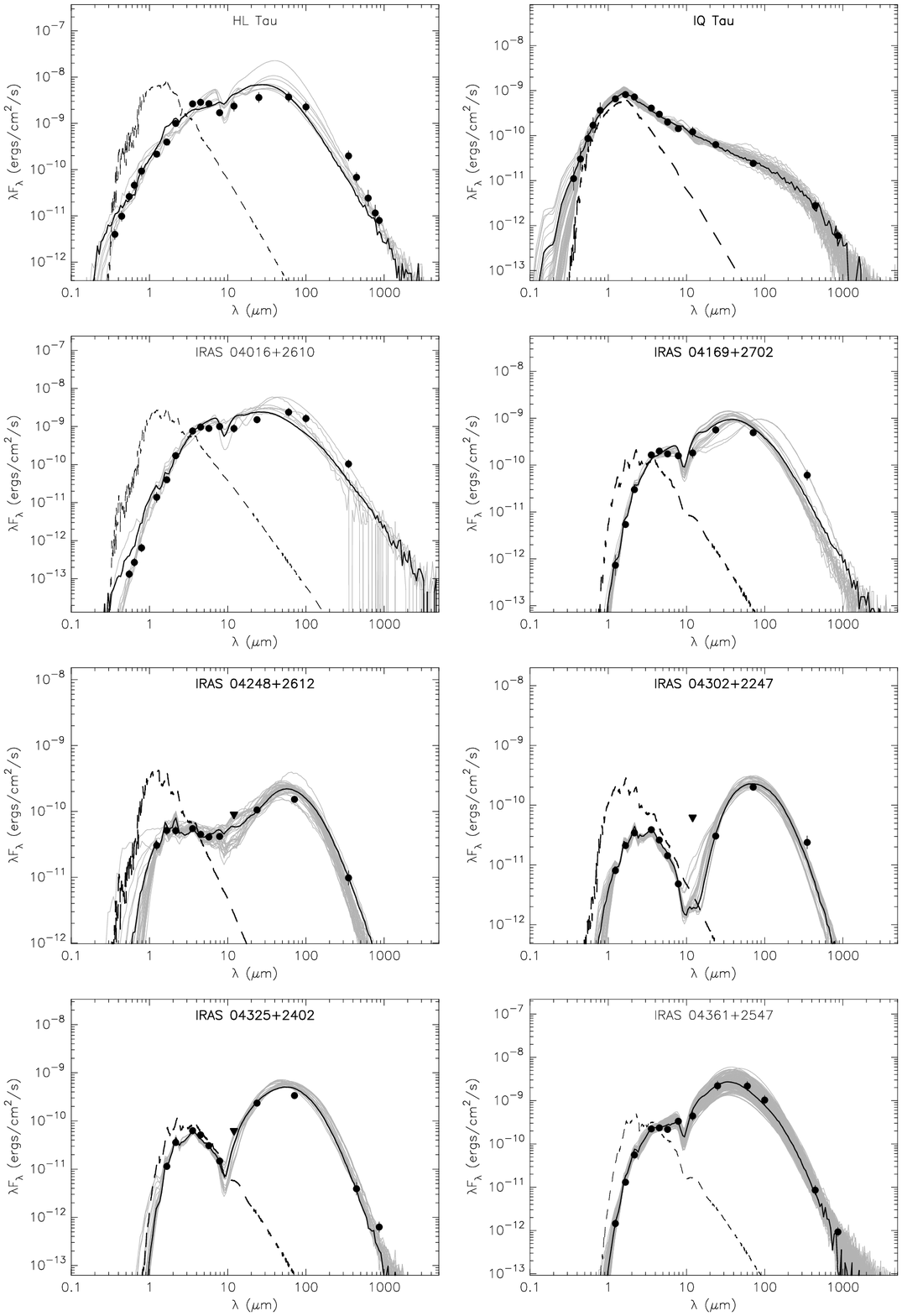}}\\[5mm]
\centerline{Fig. 1. --- Continued.}
\clearpage
{\plotone{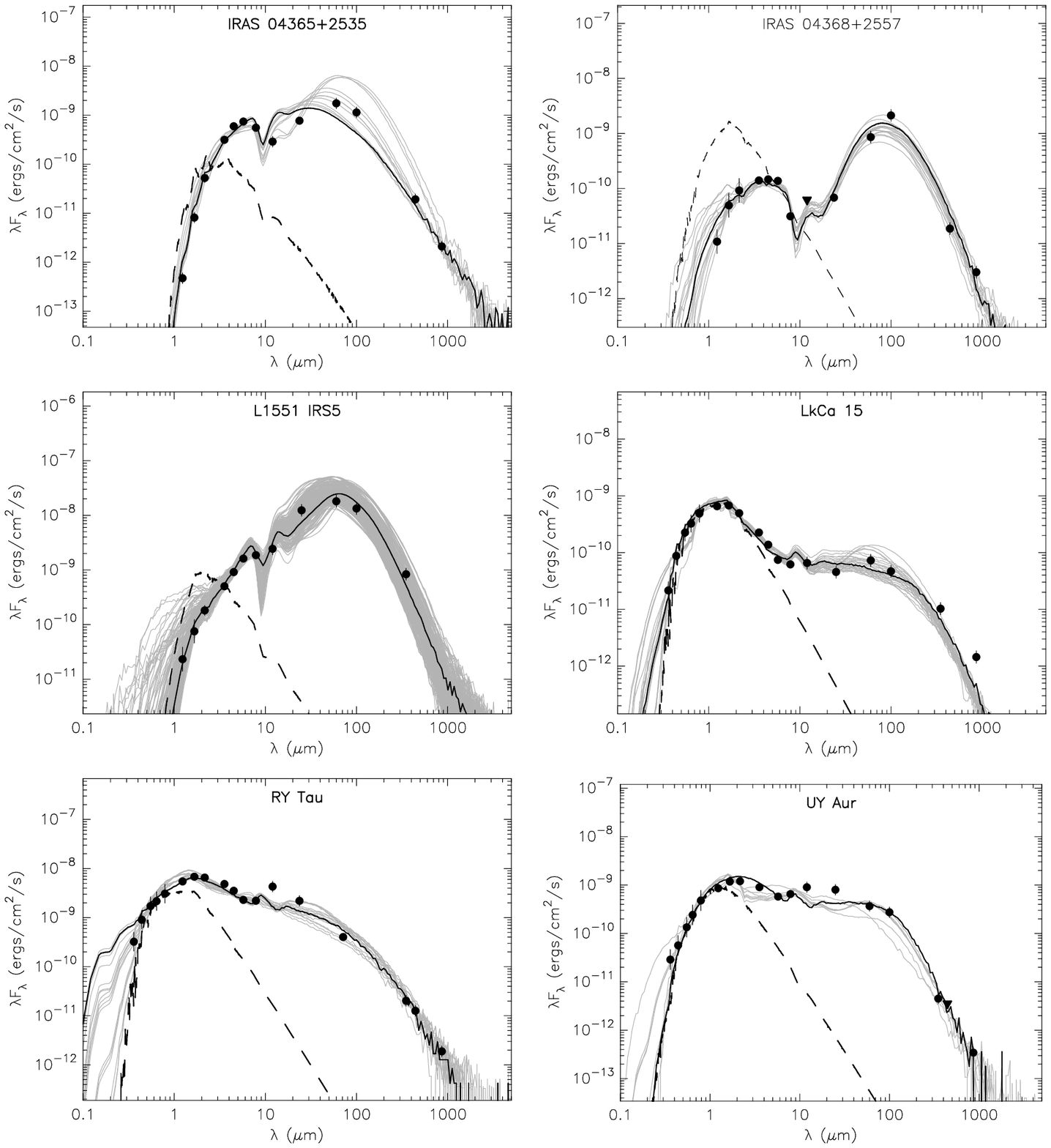}}\\[5mm]
\centerline{Fig. 1. --- Continued.}
\clearpage

\clearpage

\clearpage

\clearpage
 
\clearpage

\clearpage

%%% ENVELOPE ACCRETION RATE %%%

\clearpage

\begin{deluxetable}{lcccccccc}
\tabletypesize{\footnotesize}
\tablecolumns{5}
\tablewidth{0pt}
\tablecaption{Comparison of values of the envelope accretion rate found from SED fitting with the evolutionary stage of the object determined from resolved observations.\label{t:mdot}}
\tablehead{ & & \multicolumn{3}{c}{SED Fitting values} & \multicolumn{2}{c}{$i$ ($\mdote=0$)} & \multicolumn{2}{c}{$i$ ($\mdote>0$)}\\
Source Name & Evol. Stage & Min & Best & Max & Min & Max & Min & Max}
%\colhead{(1)} & \colhead{(2)} & \colhead{(3)} & \colhead{(4)} & \colhead{(5)}} 
%\input{tab1.tex}
\startdata
AA Tau           & Disk & 0 & 0 & 7.69$\times10^{-8}$				
 & 18 & 81 & 18 & 63	              \\ 
AB Aur           & Hae & 0 & 0 & 1.41$\times10^{-6}$				
& 32 & 63 & 32 & 57	              \\ 
BP Tau           & Disk & 0 & 5.18$\times10^{-8}$ & 7.69$\times10^{-8}$ 	
		 & 18 & 81 & 18 & 81	              \\ 
CI Tau           & Disk & 0 & 0 & 5.72$\times10^{-8}$ 				
 & 18 & 76 & 18 & 63	              \\ 
CoKu Tau 1       & Embedded/Disk & 6.88$\times10^{-8}$ & 6.88$\times10^{-8}$ &
6.88$\times10^{-8}$ 		 & \nodata & \nodata & 87 & 87        \\ 
CY Tau           & Disk & 0 & 0 & 0 						
& 18 & 81 & \nodata & \nodata        \\ 
DG Tau           & Disk & 0 & 5.48$\times10^{-9}$ & 1.11$\times10^{-7}$ 	
		 & 18 & 81 & 49 & 81	              \\ 
DG Tau B         & Embedded & 0 & 1.00$\times10^{-6}$ & 1.00$\times10^{-6}$ 	
& 87 & 87 & 18 & 87	              \\ 
DL Tau           & Disk & 0 & 0 & 0 						
& 18 & 63 & \nodata & \nodata        \\ 
DM Tau           & Disk & 7.08$\times10^{-9}$ & 6.19$\times10^{-7}$ &
8.89$\times10^{-6}$ 			 & \nodata & \nodata & 18 & 49       
\\ 
DN Tau           & Disk & 0 & 0 & 5.18$\times10^{-8 }$				
 & 18 & 81 & 18 & 63	              \\ 
DO Tau           & Disk & 0 & 7.22$\times10^{-9}$ & 7.22$\times10^{-9}$ 	
		 & 32 & 41 & 18 & 81	              \\ 
DR Tau           & Disk & 0 & 7.22$\times10^{-9}$ & 5.67$\times10^{-7}$ 	
		 & 18 & 76 & 18 & 76	              \\ 
FT Tau           & Disk & 0 & 4.73$\times10^{-8}$ & 7.21$\times10^{-6}$ 	
		 & 18 & 81 & 18 & 76	              \\ 
GG Tau           & Disk & 0 & 0 & 2.31$\times10^{-6}$ 				
 & 18 & 57 & 18 & 41	              \\ 
GM Aur           & Disk & 0 & 0 & 1.01$\times10^{-6}$				
 & 18 & 57 & 18 & 41	              \\ 
HL Tau           & Embedded & 1.87$\times10^{-6}$ & 2.13$\times10^{-6}$ &
1.59$\times10^{-5}$ 			 & \nodata & \nodata & 18 & 32       
\\ 
IQ Tau           & Disk & 0 & 0 & 5.18$\times10^{-8}$ 				
 & 18 & 81 & 18 & 81	              \\ 
IRAS 04016+2610  & Embedded & 0 & 9.28$\times10^{-7}$ & 4.86$\times10^{-6}$	
& 81 & 87 & 18 & 41	              \\ 
IRAS 04169+2702  & Embedded & 3.70$\times10^{-7}$ & 1.52$\times10^{-6}$ &
2.78$\times10^{-5}$			 & \nodata & \nodata & 32 & 76       
\\ 
IRAS 04248+2612  & Embedded & 0 & 1.73$\times10^{-6}$ & 5.10$\times10^{-6}$	
& 81 & 81 & 18 & 87	              \\ 
IRAS 04302+2247  & Embedded & 1.65$\times10^{-6}$ & 1.23$\times10^{-5}$ &
1.23$\times10^{-5}$ 			 & \nodata & \nodata & 18 & 81       
\\ 
IRAS 04325+2402  & Embedded & 1.20$\times10^{-6}$ & 1.20$\times10^{-6}$ &
1.51$\times10^{-5}$ 			 & \nodata & \nodata & 32 & 76       
\\ 
IRAS 04361+2547  & Embedded & 8.51$\times10^{-7}$ & 1.46$\times10^{-6}$ &
3.00$\times10^{-5}$ 			 & \nodata & \nodata & 18 & 81       
\\ 
IRAS 04365+2535  & Embedded & 7.16$\times10^{-7}$ & 9.28$\times10^{-7}$ &
2.64$\times10^{-5}$ 			 & \nodata & \nodata & 18 & 76       
\\ 
IRAS 04368+2557  & Embedded & 9.11$\times10^{-6}$ & 2.82$\times10^{-5}$ &
5.05$\times10^{-5}$ 			 & \nodata & \nodata & 41 & 81       
\\ 
L1551 IRS5       & Embedded & 5.54$\times10^{-6}$ & 6.47$\times10^{-5}$ &
2.96$\times10^{-4}$ 			 & \nodata & \nodata & 18 & 76       
\\ 
LkCa 15          & Disk & 0 & 0 & 3.76$\times10^{-7}$				
 & 18 & 70 & 18 & 63	              \\ 
RY Tau           & Disk & 0 & 0 & 9.52$\times10^{-8}$ 				
 & 18 & 76 & 57 & 76	              \\ 
UY Aur           & Disk & 0 & 0 & 5.88$\times10^{-7}$				
 & 70 & 70 & 57 & 81
\enddata
\tablecomments{All accretion rates are $\msol$/yr. `Embedded'
refers to sources
that are still surrounded by an infalling envelope, and `Disk' refers to
sources that are surrounded only by a circumstellar disk.
The last four columns show the ranges of inclinations $i$ for the disk-only models ($\mdote=0$) which provide a good fit as well as the range
of inclinations $i$ for the embedded models ($\mdote>0$) which provide a good fit. In the Monte-Carlo radiation transfer code (see Paper I),
photons are binned into ten viewing angles. The values quoted here correspond to the central value of the bins. The central values of the bins are, from face-on to
edge-on respectively: 18$^{\circ}$, 32$^{\circ}$, 41$^{\circ}$, 49$^{\circ}$, 57$^{\circ}$, 63$^{\circ}$, 70$^{\circ}$, 76$^{\circ}$, 81$^{\circ}$, and 87$^{\circ}$ (to the nearest degree)}
\end{deluxetable}

\clearpage

%%% STELLAR TEMPERATURE %%%

\begin{deluxetable}{lcccc}
\tabletypesize{\footnotesize}
\tablecolumns{5}
\tablewidth{0pt}
\tablecaption{Comparison of values of the central source temperatures from the literature with those found from SED fitting.\label{t:temp}}
\tablehead{ & & \multicolumn{3}{c}{SED Fitting values} \\ 
Source Name & $T_{star}$ & $T_{\rm min}$ & $T_{\rm best}$ & $T_{\rm max}$}
%\colhead{(1)} & \colhead{(2)} & \colhead{(3)} & \colhead{(4)} & \colhead{(5)}}
%\input{tab2.tex}
\startdata
AA Tau           & 4060 & 3060 & 4458 & 4859 \\ 
AB Aur           & 10500 & 4881 & 11767 & 13452 \\ 
BP Tau           & 4060 & 3138 & 3427 & 5415 \\ 
CI Tau           & 4060 & 3486 & 4255 & 4859 \\ 
CoKu Tau 1       & \nodata & 4826 & 4826 & 4826 \\ 
CY Tau           & 3720 & 3855 & 4329 & 4329 \\ 
DG Tau           & 4350$\rightarrow$5080 & 4314 & 4549 & 10722 \\ 
DG Tau B         & \nodata & 2580 & 2706 & 13618 \\ 
DL Tau           & 4060 & 4255 & 4255 & 4458 \\ 
DM Tau           & 3720 & 2762 & 2993 & 3045 \\ 
DN Tau           & 3850 & 3427 & 4314 & 4458 \\ 
DO Tau           & 3850 & 4123 & 4123 & 5209 \\ 
DR Tau           & 4060 & 3935 & 4123 & 8061 \\ 
FT Tau           & \nodata & 2783 & 3060 & 5013 \\ 
GG Tau           & 4060 & 3320 & 4430 & 4859 \\ 
GM Aur           & 4730 & 3303 & 5126 & 5126 \\ 
HL Tau           & 4060 & 2561 & 3409 & 4030 \\ 
IQ Tau           & 3785 & 3427 & 4255 & 5612 \\ 
IRAS 04016+2610  & \nodata & 2552 & 2847 & 13587 \\ 
IRAS 04169+2702  & \nodata & 2585 & 2718 & 4152 \\ 
IRAS 04248+2612  & 3580 & 2627 & 2909 & 4956 \\ 
IRAS 04302+2247  & \nodata & 2762 & 2986 & 3241 \\ 
IRAS 04325+2402  & \nodata & 2586 & 2741 & 3616 \\ 
IRAS 04361+2547  & \nodata & 2585 & 3225 & 4064 \\ 
IRAS 04365+2535  & \nodata & 2637 & 2847 & 4228 \\ 
IRAS 04368+2557  & \nodata & 2932 & 3855 & 3869 \\ 
L1551 IRS5       & 4060$\rightarrow$6030 & 3070 & 3651 & 4900 \\ 
LkCa 15          & 4350 & 3017 & 4150 & 4204 \\ 
RY Tau           & 5080 & 4604 & 4604 & 5713 \\ 
UY Aur           & 4060 & 3540 & 8090 & 8090 
\enddata
\tablecomments{All temperatures are in K.}
\tablerefs{All literature spectral types were taken from \citet{kh95}, except that for DG~Tau, which is from \citet{white04}. We use Table A5 in \citet{kh95} to convert the spectral types to temperatures.}
\end{deluxetable}

\clearpage

%%% DISK MASS %%%

\begin{deluxetable}{lccccccccc}
\tabletypesize{\footnotesize}
\tablecolumns{10}
\tablewidth{0pt}
\tablecaption{Comparison of values of the disk mass from the literature with those found from SED fitting.\label{t:mdisk}}
\tablehead{ & \multicolumn{6}{c}{Literature values} & \multicolumn{3}{c}{SED Fitting values} \\ 
Source Name & D96 & K02 & A05 & Mean & Min & Max & Min & Best & Max}
%\colhead{(1)} & \colhead{(2)} & \colhead{(3)} & \colhead{(4)} & \colhead{(5)} & \colhead{(6)} & \colhead{(7)} & \colhead{(8)} & \colhead{(9)} & \colhead{(10)}}
%\input{tab3.tex}
\startdata
AA Tau           & -1.8 & -1.6 $\pm$ 0.3 & -1.9 $\pm$ 0.1 & -1.7 & -2.0 & -1.3 & -2.1 & -1.7 & -1.5 \\ 
AB Aur           & - & \nodata & -2.4 $\pm$ 0.1 & -2.4 & -2.4 & -2.3 & -1.5 & -1.5 & -1.0 \\ 
BP Tau           & -2.9 & \nodata & -1.8 $\pm$ 0.1 & -2.0 & -2.9 & -1.7 & -3.0 & -1.9 & -1.4 \\ 
CI Tau           & -1.2 & \nodata & -1.6 $\pm$ 0.1 & -1.4 & -1.7 & -1.2 & -2.0 & -1.5 & -1.4 \\ 
CoKu Tau 1       & - & \nodata & $\sim$ -2.7 & -2.7 & -2.7 & -2.7 & -3.6 & -3.6 & -3.6 \\ 
CY Tau           & -1.8 & -1.3 $\pm$ 0.4 & $\sim$ -2.2 & -1.5 & -2.2 & -0.8 & -2.1 & -1.7 & -1.7 \\ 
DG Tau           & -1.6 & \nodata & -1.6 $\pm$ 0.1 & -1.6 & -1.7 & -1.6 & -3.7 & -1.2 & -1.2 \\ 
DG Tau B         & - & \nodata & \nodata& \nodata & \nodata & \nodata & -3.7 & -2.6 & -1.1 \\ 
DL Tau           & - & \nodata & -1.1 $\pm$ 0.1 & -1.0 & -1.2 & -1.0 & -1.7 & -1.5 & -1.5 \\ 
DM Tau           & -1.6 & -1.7 $\pm$ 0.3 & -1.6 $\pm$ 0.1 & -1.6 & -2.0 & -1.4 & -3.1 & -2.4 & -2.2 \\ 
DN Tau           & -1.6 & -9.0 $\pm$ 2.5 & -1.5 $\pm$ 0.1 & -1.6 & \nodata & -0.8 & -2.1 & -1.5 & -1.5 \\ 
DO Tau           & -1.7 & -2.7 $\pm$ 0.5 & -2.2 $\pm$ 0.1 & -2.0 & \nodata & -1.7 & -1.9 & -1.5 & -1.5 \\ 
DR Tau           & - & -2.3 $\pm$ 0.3 & -1.7 $\pm$ 0.1 & -1.9 & -2.7 & -1.7 & -2.0 & -1.5 & -1.2 \\ 
FT Tau           & -1.9 & \nodata & -1.9 $\pm$ 0.1 & -1.9 & -1.9 & -1.8 & -3.0 & -2.1 & -1.5 \\ 
GG Tau           & - & \nodata & -0.7 $\pm$ 0.2 & -0.6 & -0.8 & -0.5 & -1.9 & -1.5 & -1.5 \\ 
GM Aur           & -1.4 & -1.5 $\pm$ 0.3 & -1.6 $\pm$ 0.1 & -1.5 & -1.9 & -1.2 & -2.6 & -1.5 & -1.4 \\ 
HL Tau           & - & -1.3 $\pm$ 0.1 & -1.2 $\pm$ 0.1 & -1.2 & -1.4 & -1.1 & -1.9 & -1.5 & -1.1 \\ 
IQ Tau           & - & -1.4 $\pm$ 0.2 & -1.7 $\pm$ 0.1 & -1.5 & -1.7 & -1.2 & -2.3 & -1.5 & -1.4 \\ 
IRAS 04016+2610  & - & \nodata & $\sim$ -1.7 & -1.7 & -1.7 & -1.7 & -2.2 & -1.8 & -0.9 \\ 
IRAS 04169+2702  & - & \nodata & $\sim$ -1.5 & -1.5 & -1.5 & -1.5 & -3.0 & -1.8 & -1.3 \\ 
IRAS 04248+2612  & - & \nodata & $\sim$ -2.3 & -2.3 & -2.3 & -2.3 & -4.3 & -2.2 & -1.4 \\ 
IRAS 04302+2247  & - & \nodata & $\sim$ -1.5 & -1.5 & -1.5 & -1.5 & -5.5 & -2.7 & -2.2 \\ 
IRAS 04325+2402  & - & \nodata & $\sim$ -2.1 & -2.1 & -2.1 & -2.1 & -4.6 & -2.8 & -2.5 \\ 
IRAS 04361+2547  & - & \nodata & $\sim$ -2.0 & -2.0 & -2.0 & -2.0 & -4.6 & -3.6 & -1.5 \\ 
IRAS 04365+2535  & - & \nodata & $\sim$ -1.5 & -1.5 & -1.5 & -1.5 & -3.4 & -1.8 & -1.2 \\ 
IRAS 04368+2557  & - & \nodata & $\sim$ -1.4 & -1.4 & -1.4 & -1.4 & -4.2 & -3.6 & -1.8 \\ 
L1551 IRS5       & - & \nodata & $\sim$ -0.3 & -0.3 & -0.3 & -0.3 & -3.5 & -1.5 & -0.9 \\ 
LkCa 15          & - & -1.4 $\pm$ 0.4 & -1.3 $\pm$ 0.1 & -1.3 & -2.1 & -1.0 & -2.1 & -2.0 & -1.7 \\ 
RY Tau           & - & -3.4 $\pm$ 1.0 & -1.8 $\pm$ 0.1 & -1.9 & \nodata & -1.7 & -2.4 & -1.5 & -1.0 \\ 
UY Aur           & - & \nodata & -2.8 $\pm$ 0.1 & -2.7 & -2.8 & -2.7 & -2.9 & -1.4 & -1.4  
\enddata
\tablecomments{All disk masses are in $\log{[\msol]}$.}
\tablerefs{D96: \citet{dutrey96}, K02: \citet{kitamura02}, A05: \citet{andrews05}.}
\end{deluxetable}

\clearpage

%%% DISK ACCRETION RATE %%%

\begin{deluxetable}{lcccccccccc}
\tabletypesize{\footnotesize}
\tablecolumns{11}
\tablewidth{0pt}
\tablecaption{Comparison of values of the disk accretion rate from the literature with those found from SED fitting.\label{t:mdotdisk}}
\tablehead{ & \multicolumn{7}{c}{Literature values} & \multicolumn{3}{c}{SED Fitting values}\\
Source Name & V93 & H95 & H98 & M05 & Mean & Min & Max & Min & Best & Max}
%\colhead{(1)} & \colhead{(2)} & \colhead{(3)} & \colhead{(4)} & \colhead{(5)} & \colhead{(6)} & \colhead{(7)} & \colhead{(8)} & \colhead{(9)} & \colhead{(10)} & \colhead{(11)}} 
%\input{tab4.tex}
\startdata
AA Tau           & -8.15 & -6.90 & -8.48 & -8.56 & -8.02 & -8.56 & -6.90 & -7.96 & -7.16 & -6.8 \\ 
AB Aur           & \nodata & \nodata & \nodata & \nodata & \nodata & \nodata & \nodata & -7.13 & -6.92 & -5.88 \\ 
BP Tau           & -7.61 & -6.80 & -7.54 & -7.71 & -7.42 & -7.71 & -6.80 & -8.26 & -7.19 & -6.71 \\ 
CI Tau           & -7.83 & -6.80 & -7.19 & \nodata & -7.27 & -7.83 & -6.80 & -7.31 & -6.93 & -6.58 \\ 
CoKu Tau 1       & \nodata & \nodata & \nodata & \nodata & \nodata & \nodata & \nodata & -10.08 & -10.08 & -10.08 \\ 
CY Tau           & \nodata & -8.20 & -8.12 & -8.16 & -8.16 & -8.20 & -8.12 & -8.52 & -6.95 & -6.95 \\ 
DG Tau           & \nodata & -5.70 & \nodata & -6.93 & -6.32 & -6.93 & -5.70 & -11.2 & -6.07 & -6.07 \\ 
DG Tau B         & \nodata & \nodata & \nodata & \nodata & \nodata & \nodata & \nodata & -9.98 & -6.98 & -5.64 \\ 
DL Tau           & -7.63 & -6.70 & \nodata & \nodata & -7.17 & -7.63 & -6.70 & -7.16 & -6.93 & -6.93 \\ 
DM Tau           & -8.54 & \nodata & -7.95 & \nodata & -8.24 & -8.54 & -7.95 & -7.93 & -7.47 & -7.33 \\ 
DN Tau           & -8.89 & -7.50 & -8.46 & -8.72 & -8.39 & -8.89 & -7.50 & -8.26 & -7.56 & -6.93 \\ 
DO Tau           & -7.22 & -5.60 & -6.84 & -7.15 & -6.70 & -7.22 & -5.60 & -6.82 & -6.17 & -6.17 \\ 
DR Tau           & \nodata & -5.10 & \nodata & -6.66 & -5.88 & -6.66 & -5.10 & -8.16 & -6.17 & -6 \\ 
FT Tau           & \nodata & \nodata & \nodata & \nodata & \nodata & \nodata & \nodata & -8.56 & -7.29 & -6.79 \\ 
GG Tau           & -7.52 & -6.70 & -7.76 & \nodata & -7.33 & -7.76 & -6.70 & -7.88 & -7.16 & -6.58 \\ 
GM Aur           & -8.13 & -7.60 & -8.02 & -8.11 & -7.97 & -8.13 & -7.60 & -8.5 & -7.37 & -7.37 \\ 
HL Tau           & \nodata & \nodata & \nodata & \nodata & \nodata & \nodata & \nodata & -6.53 & -5.99 & -5.29 \\ 
IQ Tau           & -7.74 & \nodata & -7.55 & \nodata & -7.65 & -7.74 & -7.55 & -8.26 & -6.93 & -6.71 \\ 
IRAS 04016+2610  & \nodata & \nodata & \nodata & \nodata & \nodata & \nodata & \nodata & -7.22 & -6.26 & -5.41 \\ 
IRAS 04169+2702  & \nodata & \nodata & \nodata & \nodata & \nodata & \nodata & \nodata & -7.76 & -6.73 & -6.12 \\ 
IRAS 04248+2612  & \nodata & \nodata & \nodata & \nodata & \nodata & \nodata & \nodata & -10.41 & -7.57 & -6.89 \\ 
IRAS 04302+2247  & \nodata & \nodata & \nodata & \nodata & \nodata & \nodata & \nodata & -11.2 & -9.3 & -7.33 \\ 
IRAS 04325+2402  & \nodata & \nodata & \nodata & \nodata & \nodata & \nodata & \nodata & -11.1 & -8.88 & -7.28 \\ 
IRAS 04361+2547  & \nodata & \nodata & \nodata & \nodata & \nodata & \nodata & \nodata & -11.97 & -8.56 & -5.55 \\ 
IRAS 04365+2535  & \nodata & \nodata & \nodata & \nodata & \nodata & \nodata & \nodata & -8.6 & -6.26 & -5.68 \\ 
IRAS 04368+2557  & \nodata & \nodata & \nodata & \nodata & \nodata & \nodata & \nodata & -11.01 & -11.01 & -7.3 \\ 
L1551 IRS5       & \nodata & \nodata & \nodata & \nodata & \nodata & \nodata & \nodata & -9.29 & -5.34 & -4.93 \\ 
LkCa 15          & \nodata & \nodata & -8.87 & \nodata & -8.87 & -8.87 & -8.87 & -8.55 & -8.33 & -7.16 \\ 
RY Tau           & \nodata & -7.60 & \nodata & \nodata & -7.60 & -7.60 & -7.60 & -6.61 & -5.98 & -5.92 \\ 
UY Aur           & -8.31 & -6.60 & -7.18 & \nodata & -7.36 & -8.31 & -6.60 & -8.94 & -8.57 & -6.59  
\enddata
\tablecomments{All disk accretion rates are in $\log{[\msol{\rm /yr}]}$.}
\tablerefs{V93: \citet{valenti93}, H95: \citet{hartigan95}, H98: \citet{hartmann98}, M05: \citet{mohanty05}.}

\end{deluxetable}

%%% ADDITIONAL PARAMETERS %%%

\begin{deluxetable}{lcccccccc}
\tabletypesize{\footnotesize}
\tablecolumns{11}
\tablewidth{0pt}
\tablecaption{Range of parameter values providing a good fit from SED fitting for additional disk parameters.\label{t:addpar}}
\tablehead{Source Name & $\rmind$ & $\rmind$ & $\rmind$ & $\rmind$ & $\rmaxd$ & $\rmaxd$ & $h$(100\,AU) & $h$(100\,AU) \\
& Min & Max & Min & Max & Min & Max & Min & Max \\
& ($\rsub$) & ($\rsub$) & (AU) & (AU) & (AU) & (AU) & (AU) & (AU)}
\startdata
AA Tau           & 1.0 & 6.4 & 0.04 & 0.51 & 62.1 & 168.9 & 2.33 & 6.67\\ 
AB Aur           & 1.0 & 2.7 & 0.48 & 1.19 & 83.1 & 170.7 & 2.90 & 7.28\\ 
BP Tau           & 1.0 & 6.4 & 0.05 & 0.51 & 19.3 & 585.8 & 1.17 & 9.26\\ 
CI Tau           & 1.0 & 6.8 & 0.08 & 0.62 & 77.8 & 198.7 & 1.17 & 5.51\\ 
CoKu Tau 1       & 1.0 & 1.0 & 0.23 & 0.23 & 76.7 & 76.7 & 5.02 & 5.02\\ 
CY Tau           & 1.0 & 1.0 & 0.04 & 0.11 & 97.5 & 198.7 & 1.17 & 3.79\\ 
DG Tau           & 1.0 & 6.1 & 0.21 & 2.50 & 78.8 & 2540.7 & 2.69 & 6.81\\ 
DG Tau B         & 1.0 & 9.5 & 0.07 & 6.25 & 2.3 & 893.7 & 2.76 & 10.37\\ 
DL Tau           & 1.0 & 1.0 & 0.10 & 0.11 & 101.3 & 198.7 & 1.17 & 3.16\\ 
DM Tau           & 130.2 & 455.1 & 3.37 & 19.68 & 50.4 & 178.3 & 3.53 & 11.64\\ 
DN Tau           & 1.0 & 6.4 & 0.06 & 0.51 & 49.6 & 163.1 & 2.33 & 5.77\\ 
DO Tau           & 1.0 & 1.5 & 0.16 & 0.26 & 103.7 & 190.6 & 2.42 & 2.91\\ 
DR Tau           & 1.0 & 5.8 & 0.21 & 1.39 & 72.4 & 158.5 & 2.26 & 7.61\\ 
FT Tau           & 1.0 & 12.1 & 0.04 & 1.01 & 24.8 & 489.3 & 1.17 & 17.42\\ 
GG Tau           & 1.0 & 6.8 & 0.06 & 0.62 & 55.0 & 133.0 & 3.14 & 12.47\\ 
GM Aur           & 12.1 & 265.8 & 1.01 & 23.93 & 69.3 & 352.2 & 2.30 & 5.84\\ 
HL Tau           & 1.0 & 2.9 & 0.17 & 0.43 & 2.2 & 39.0 & 4.80 & 19.30\\ 
IQ Tau           & 1.0 & 6.4 & 0.07 & 0.51 & 49.6 & 415.8 & 1.17 & 5.77\\ 
IRAS 04016+2610  & 1.0 & 2.9 & 0.09 & 2.46 & 2.2 & 3651.1 & 3.35 & 30.16\\ 
IRAS 04169+2702  & 1.0 & 3.9 & 0.07 & 0.42 & 3.6 & 150.7 & 5.28 & 27.90\\ 
IRAS 04248+2612  & 1.0 & 48.9 & 0.03 & 2.20 & 27.0 & 211.5 & 2.74 & 13.97\\ 
IRAS 04302+2247  & 77.9 & 1700.5 & 3.25 & 58.85 & 36.2 & 1215.0 & 3.31 & 10.12\\ 
IRAS 04325+2402  & 95.6 & 497.5 & 4.29 & 21.55 & 22.8 & 393.5 & 3.69 & 14.99\\ 
IRAS 04361+2547  & 1.0 & 11.3 & 0.07 & 1.63 & 2.2 & 1296.0 & 3.43 & 24.36\\ 
IRAS 04365+2535  & 1.0 & 9.3 & 0.09 & 1.60 & 2.7 & 176.7 & 5.11 & 15.54\\ 
IRAS 04368+2557  & 1.0 & 2.5 & 0.05 & 0.24 & 33.8 & 1303.3 & 3.18 & 9.67\\ 
L1551 IRS5       & 1.0 & 30.0 & 0.17 & 12.10 & 1.2 & 731.2 & 2.81 & 21.12\cr
LkCa 15          & 1.0 & 2.7 & 0.04 & 0.12 & 62.1 & 195.5 & 4.42 & 9.84\cr
RY Tau           & 1.0 & 4.7 & 0.29 & 1.38 & 37.9 & 124.1 & 1.92 & 5.76\cr
UY Aur           & 1.0 & 10.1 & 0.27 & 1.94 & 51.5 & 1159.7 & 3.47 & 10.44 
\enddata
\end{deluxetable}

%%% ADDITIONAL PARAMETERS 2 %%%

\begin{deluxetable}{lcccc}
\tabletypesize{\footnotesize}
\tablecolumns{11}
\tablewidth{0pt}
\tablecaption{Range of parameter values providing a good fit from SED fitting for the stellar mass and total bolometric luminosity.\label{t:addpar2}}
\tablehead{Source Name & $\mstar$ & $\mstar$ & $\lstar$ & $\lstar$ \\
& Min & Max & Min & Max \\
& ($\msol$) & ($\msol$) & ($\lsol$) & ($\lsol$)}
\startdata
AA Tau           & 0.20 & 1.74 & 0.44 & 3.32\\ 
AB Aur           & 2.57 & 3.62 & 42.64 & 181.47\\ 
BP Tau           & 0.22 & 1.74 & 0.59 & 4.48\\ 
CI Tau           & 0.35 & 1.78 & 1.60 & 5.12\\ 
CoKu Tau 1       & 2.68 & 2.68 & 11.44 & 11.44\\ 
CY Tau           & 0.57 & 1.08 & 0.40 & 3.01\\ 
DG Tau           & 1.10 & 3.07 & 9.63 & 42.02\\ 
DG Tau B         & 0.11 & 3.70 & 0.67 & 249.65\\ 
DL Tau           & 0.99 & 1.29 & 2.53 & 3.01\\ 
DM Tau           & 0.10 & 0.20 & 0.16 & 0.44\\ 
DN Tau           & 0.33 & 1.29 & 0.82 & 2.70\\ 
DO Tau           & 0.80 & 1.49 & 5.40 & 7.59\\ 
DR Tau           & 0.63 & 2.13 & 7.53 & 14.36\\ 
FT Tau           & 0.11 & 2.03 & 0.36 & 7.01\\ 
GG Tau           & 0.28 & 1.63 & 0.82 & 3.01\\ 
GM Aur           & 0.28 & 1.62 & 0.65 & 3.16\\ 
HL Tau           & 0.10 & 0.91 & 3.46 & 15.05\\ 
IQ Tau           & 0.33 & 1.94 & 1.04 & 5.16\\ 
IRAS 04016+2610  & 0.10 & 3.73 & 2.00 & 249.21\\ 
IRAS 04169+2702  & 0.11 & 0.90 & 0.73 & 5.87\\ 
IRAS 04248+2612  & 0.10 & 2.84 & 0.27 & 11.30\\ 
IRAS 04302+2247  & 0.10 & 0.25 & 0.15 & 0.67\\ 
IRAS 04325+2402  & 0.10 & 0.41 & 0.46 & 1.54\\ 
IRAS 04361+2547  & 0.11 & 0.85 & 1.20 & 9.37\\ 
IRAS 04365+2535  & 0.11 & 1.25 & 2.00 & 14.43\\ 
IRAS 04368+2557  & 0.15 & 0.59 & 0.64 & 3.84\\ 
L1551 IRS5       & 0.21 & 4.70 & 6.88 & 72.40\\ 
LkCa 15          & 0.18 & 0.91 & 0.42 & 1.39\\ 
RY Tau           & 1.64 & 3.34 & 18.75 & 40.65\\ 
UY Aur           & 0.37 & 2.04 & 2.11 & 14.22 
\enddata
\end{deluxetable}

\clearpage

\begin{deluxetable}{lll}
\tablewidth{0pt}
\tablecaption{The main parameters for the best-fitting models to IRAS~04368+2557\label{t:04368tab}}
\tablehead{\colhead{Parameter} & \colhead{IRAC data only} & \colhead{Full SED}}
\startdata
Stellar Mass            &    4.07\,$\msol$ & 1.46\,$\msol$	\\
Stellar Radius           &  21.57\,$\rsol$ & 8.21\,$\rsol$	\\
Stellar Temperature      &   4360\,K & 4260\,K	\\
Envelope accretion rate  &   2.63$\times10^{-4}$\,$\msol$/yr &  1.37$\times10^{-4}$\,$\msol$/yr	\\
Envelope outer radius     &  9120\,AU & 16200\,AU	\\
Cavity opening angle      &  16$^{\circ}$ & 43$^{\circ}$	\\
Viewing angle	          &  75$^{\circ}$ & 81$^{\circ}$	\\
Bolometric Luminosity     &  155\,$\lsol$ & 20\,$\lsol$
\enddata
\tablecomments{These are the main parameters of the best-fit model SEDs to the SED of IRAS~04368+2557, first fitting  only the IRAC data simulaneously in six apertures, then fitting the full SED including the multi-aperture IRAC data.} 
\end{deluxetable}

%%%%%%%%%%%%%%%%%% DATA %%%%%%%%%%%%%%%%%%

\clearpage

\appendix

\section{The convolution of model SEDs with broadband filters}

\label{a:conv}

This appendix describes the exact procedure used to obtain monochromatic fluxes through broadband filters for our models.\\

In the following, we define the true spectrum of a source or model SED to be $\fnua$.
In general, when a broadband flux is measured through a filter, we have no knowledge of the true underlying spectrum, only the integrated flux over the filter.
Therefore, to quote a monochromatic flux $\fnuqz$ at a frequency $\nu_0$, one usually makes an assumption about the spectrum of the source.
We call this spectrum $\fnuq$.
Note the difference between $\fnuq$ and $\fnuqz$: $\fnuqz$ is the value of $\fnuq$ at $\nu_0$.\\

We make the same assumption for our model SEDs as is made for the data taken in the different filters.
For example, fluxes from the IRAC pipeline are quoted using $\fnuq\propto1/\nu$.
That is to say, what would the flux at $\nu_0$ be, if the spectrum of the source was proportional to $1/\nu$, such that the observed integrated flux was identical to what is actually observed?
Since we are making the same assumptions as used for the observed fluxes, a direct comparison between the quoted model fluxes and the quoted observed fluxes can be made without the need for any color-correction.

\subsection{Spitzer - IRAC}

The IRAC monochromatic fluxes assume $\fnuq\propto1/\nu$, or $\nu\fnuq\propto{\rm const}$ \citep{reach05}.\\

The total electron ``count'' detected through the filter with response $R(\nu)$ (in $e^{-}/{\rm photon}$) is

\begin{equation}
E=\int\frac{\fnua}{h\nu}~R(\nu)~d\nu=\int\frac{\fnuq}{h\nu}~R(\nu)~d\nu.
\end{equation}

Now $\nu\fnuq\propto{\rm const}\equiv\nu_{0}\fnuqz$, so

\begin{equation}
\nu_{0}\fnuqz\int\frac{1}{h\nu^2}~R(\nu)~d\nu=\int\frac{\fnua}{h\nu}~R(\nu)~d\nu
\end{equation}

which after rearranging gives

\begin{equation}
\fnuqz=\frac{\int \fnua\left(\nu_0/\nu\right)~R(\nu)~d\nu}{\int\left(\nu_0/\nu\right)^2~R(\nu)~d\nu}
\end{equation}

The values of $R(\nu)$ are taken from the Spitzer Science Center website\footnote{http://ssc.spitzer.caltech.edu/irac/spectral\_response.html}. The values of $\nu_0$ are given by $\nu_0=c/\lambda_0$ where $\lambda_0=3.550,~4.493,~5.731,~{\rm and}~7.872\mu m$ are the nominal wavelengths for IRAC \citep{reach05}.

\subsection{Spitzer - MIPS}

The MIPS monochromatic fluxes assume a T=$10,000$K blackbody spectrum, i.e very close to $\fnuq\propto\nu^2$, or $\fnuq/\nu^2\propto{\rm const}$ (MIPS data handbook\footnote{http://ssc.spitzer.caltech.edu/mips/dh/}).\\

The total electron ``count'' through the filter with response $R_{\nu}(\nu)\equiv R(\nu)/\nu$ (in $e^{-}/{\rm unit~energy}$) is

\begin{equation}
E=\int\fnua~R_{\nu}(\nu)~d\nu=\int\fnuq~R_{\nu}(\nu)~d\nu.
\end{equation}

Now $\fnuq/\nu^2\propto{\rm const}\equiv\fnuqz/\nu_0^2$, so

\begin{equation}
\fnuqz/\nu_{0}^2\int\nu^2~R_{\nu}(\nu)~d\nu=\int\fnua~R_{\nu}(\nu)~d\nu
\end{equation}

which after rearranging gives

\begin{equation}
\fnuqz=\frac{\int \fnua~R_{\nu}(\nu)~d\nu}{\int\left(\nu/\nu_0\right)^2~R_{\nu}(\nu)~d\nu}
\end{equation}

The values of $R_\nu(\nu)$ are taken from the Spitzer Science Center website\footnote{http://ssc.spitzer.caltech.edu/mips/spectral\_response.html}. The values of $\nu_0$ are given by $\nu_0=c/\lambda_0$ where $\lambda_0=23.68,~71.42,~{\rm and}~155.9\mu m$ are the effective wavelengths for MIPS.

\subsection{IRAS}

The IRAS monochromatic fluxes use $\fnuq\propto1/\nu$, or $\nu\fnuq\propto{\rm const}$ (IRAS Explanatory Supplement - Section VI.C.3\footnote{http://irsa.ipac.caltech.edu/IRASdocs/exp.sup/ch6/C3.html}). In the same way as for IRAC, we have

\begin{equation}
\fnuqz=\frac{\int \fnua\left(\nu_0/\nu\right)~R(\nu)~d\nu}{\int\left(\nu_0/\nu\right)^2~R(\nu)~d\nu}
\end{equation}

However, the relative system response listed in the IRAS documentation is in electrons per unit energy (as for MIPS). Therefore, writing $\rnu(\nu)\equiv R(\nu)/\nu$, we get

\begin{equation}
\fnuqz=\frac{\int \fnua~\rnu(\nu)~d\nu}{\int\left(\nu_0/\nu\right)~\rnu(\nu)~d\nu}
\end{equation}

The values of $\rnu(\nu)$ are taken from the IRAS documentation (The `Relative System Response' in the IRAS Explanatory Supplement - Table II.C.5\footnote{http://irsa.ipac.caltech.edu/IRASdocs/exp.sup/ch2/tabC5.html}). The values of $\nu_0$ are given by $\nu_0=c/\lambda_0$ where $\lambda_0=12,~25,~60,~{\rm and}~100\mu m$.

\subsection{UBVRI photometry}

The UBVRI observations from \citet{kh95} are originally from \citet{herbst94}, and were made in the Johnson/Cousins UBVRI system.
The transmission curves for these bands were taken from \citet{bessell90}.
Since flux densities are not commonly used at optical wavelengths, it was not clear what spectrum to assume in order to derive the monochromatic fluxes (although the calibration is usually done using the spectrum of Vega).
However, we have found that the differences arising from various assumptions do not change the resulting fluxes by more than a few \%, which is much smaller than the $\pm25\%$ uncertainties we imposed on the observed UBVRI fluxes.
Therefore, the choice of the assumption is unimportant for this work.
We choose to assume a flat spectrum, i.e. $\fnuq\propto1/\nu$, or $\nu\fnuq\propto{\rm const}$.
Therefore, as for IRAC, we have:

\begin{equation}
\fnuqz=\frac{\int \fnua\left(\nu_0/\nu\right)~R(\nu)~d\nu}{\int\left(\nu_0/\nu\right)^2~R(\nu)~d\nu}
\end{equation}

The values of $\nu_0$ are given by $\nu_0=c/\lambda_0$ where $\lambda_0=0.36,~0.44,~0.55,~0.64,~{\rm and}~0.79\mu m$.

\subsection{JHK/2MASS}

For all the JHK fluxes we computed the monochromatic fluxes using the method for the 2MASS all-sky survey \citep{skrutskie06}.
The 2MASS isophotal fluxes are computed using the relative system response $\rnu$ from \citet{cohen_2mass}, and are given by:

\begin{equation}
\fnuqiso=\frac{\int \fnua~\rnu(\nu)~d\nu}{\Delta\nu_{\rm iso}}
\end{equation}

The isophotal bandwidths $\Delta\nu_{\rm iso}$ for the three bands are listed in \citet{cohen_2mass}. The isophotal wavelengths are $\lambda_0^{\rm iso}=1.235,~1.662,~{\rm and}~2.159\mu m$.

\subsection{SHARC~II, SCUBA, and CSO observations}

The SHARC~II and SCUBA instruments are calibrated on planets, whose radiation follows the Rayleigh-Jeans tail of a blackbody curve at sub-mm wavelengths.
Therefore, we take $\fnuq\propto\nu^2$, or $\fnuq/\nu^2\propto{\rm const}$.\\

As for MIPS, we have

\begin{equation}
\fnuqz=\frac{\int \fnua~R_{\nu}(\nu)~d\nu}{\int\left(\nu/\nu_0\right)^2~R_{\nu}(\nu)~d\nu}
\end{equation}

The values of $R_\nu(\nu)$ are taken from \citet{dowell03} for the SHARC~II observations, and the SCUBA website\footnote{http://www.jach.hawaii.edu/JCMT/continuum/background/background.html} for the SCUBA $450$WB and $850$WB observations.
For the CSO observations we used gaussians centered at 624 and 769\microns with FWHM 67 and 190\,\microns respectively.
We convolved the SHARC~II and CSO filters with the atmospheric transmission curve used in \citet{dowell03}, and the SCUBA filters with the atmospheric transmission curve given with the filter profiles on the SCUBA website.\\

The values of $\nu_0$ are given by $\nu_0=c/\lambda_0$ where $\lambda_0=350,~443,~{\rm and}~863\mu m$ for SHARC~II 350\,\microns and SCUBA $450$WB and $850$WB respectively, and $\lambda_0=624~{\rm and}~769\mu m$ for the CSO observations.

\clearpage

\section{The data for the 30 Taurus-Auriga sources}

\label{a:data}

\begin{deluxetable}{lrrrrrrrrr}

\tabletypesize{\tiny}
\tablecolumns{10}
\tablewidth{0pt}
\tablecaption{Optical and Near-IR data for the 30 Taurus-Auriga sources.\label{t:seds_1}}

\tablehead{Source Name & U & B & V & R & I & J & H & K & References}

\startdata
AA Tau                         &     3.60 &     8.97 &    26.24 &    48.11 &   104.42 &   271.95 &   389.67 &   404.70& 1,5 \\
 & $\pm$     1.61 & $\pm$     3.62 & $\pm$    10.20 & $\pm$    18.80 & $\pm$    40.92 & $\pm$     7.76 & $\pm$    13.28 & $\pm$    14.53& \\
AB Aur                         &  2361.72 &  5547.50 &  5723.75 &  4837.74 &  4797.35 &  6731.14 &  9671.64 & 13549.72& 1,5 \\
 & $\pm$   106.45 & $\pm$   144.47 & $\pm$   105.42 & $\pm$   125.98 & $\pm$   152.98 & $\pm$   111.57 & $\pm$   178.13 & $\pm$   199.64& \\
BP Tau                         &     9.87 &    20.89 &    51.93 &    88.44 &   175.02 &   366.85 &   546.39 &   547.94& 1,5 \\
 & $\pm$     2.01 & $\pm$     2.83 & $\pm$     4.76 & $\pm$     8.73 & $\pm$    17.93 & $\pm$    13.18 & $\pm$    15.60 & $\pm$    15.64& \\
CI Tau                         &     3.57 &     7.05 &    18.95 &    38.77 &    95.68 &   260.91 &   438.43 &   504.35& 1,5 \\
 & $\pm$     1.29 & $\pm$     2.07 & $\pm$     3.60 & $\pm$     7.36 & $\pm$    18.17 & $\pm$     6.73 & $\pm$    17.36 & $\pm$    13.93& \\
CoKu Tau 1                     & \nodata  & \nodata  & \nodata  &     1.11 &     3.41 &    11.38 &    25.98 &    27.18& 3,5 \\
 & \nodata  & \nodata  & \nodata  & $\pm$     0.00 & $\pm$     0.00 & $\pm$     0.32 & $\pm$     0.75 & $\pm$     0.61& \\
CY Tau                         &     3.01 &     5.38 &    17.20 &    33.92 &    89.30 &   197.01 &   268.35 &   249.07& 1,5 \\
 & $\pm$     0.64 & $\pm$     0.96 & $\pm$     1.58 & $\pm$     3.17 & $\pm$     8.70 & $\pm$     5.81 & $\pm$     9.64 & $\pm$     7.34& \\
DG Tau                         &    10.93 &    18.04 &    42.55 &    77.27 &   169.33 &   532.22 &   834.64 &  1064.46& 1,5 \\
 & $\pm$     3.60 & $\pm$     5.79 & $\pm$    13.37 & $\pm$    24.42 & $\pm$    53.97 & $\pm$    53.22 & $\pm$    83.46 & $\pm$   106.45& \\
DG Tau B                       & \nodata  & \nodata  & \nodata  & \nodata  & \nodata  &     1.95 &     5.91 &    14.24& 5 \\
 & \nodata  & \nodata  & \nodata  & \nodata  & \nodata  & $\pm$     0.19 & $\pm$     0.59 & $\pm$     1.42& \\
DL Tau                         &     5.14 &     8.41 &    21.80 &    55.52 &   120.44 &   229.13 &   352.13 &   438.06& 1,5 \\
 & $\pm$     1.29 & $\pm$     2.00 & $\pm$     4.33 & $\pm$    14.65 & $\pm$    31.79 & $\pm$     7.39 & $\pm$    10.70 & $\pm$    12.91& \\
DM Tau                         &     2.57 &     3.99 &     9.54 &    20.30 &    52.47 &   106.09 &   128.20 &   103.83& 1,5 \\
 & $\pm$     0.69 & $\pm$     0.80 & $\pm$     1.56 & $\pm$     3.62 & $\pm$     9.38 & $\pm$     3.03 & $\pm$     3.42 & $\pm$     3.16& \\
DN Tau                         &     2.96 &    12.29 &    41.19 &    74.77 &   168.34 &   345.53 &   474.13 &   415.27& 1,5 \\
 & $\pm$     0.66 & $\pm$     1.06 & $\pm$     3.03 & $\pm$     5.66 & $\pm$    13.56 & $\pm$    17.82 & $\pm$    21.83 & $\pm$    12.24& \\
DO Tau                         &     1.91 &     3.84 &    12.75 &    35.84 &   101.15 &   261.39 &   520.84 &   801.55& 1,5 \\
 & $\pm$     1.01 & $\pm$     1.68 & $\pm$     4.87 & $\pm$    15.76 & $\pm$    45.33 & $\pm$     2.65 & $\pm$    18.23 & $\pm$    25.10& \\
DR Tau                         &    44.27 &    56.20 &    98.50 &   140.72 &   254.35 &   476.97 &   761.20 &  1192.15& 1,5 \\
 & $\pm$    21.70 & $\pm$    26.95 & $\pm$    46.64 & $\pm$    66.66 & $\pm$   121.11 & $\pm$    16.69 & $\pm$    41.36 & $\pm$    40.62& \\
FT Tau                         & \nodata  &     2.32 &     5.03 &    22.56 &    75.30 &   143.12 &   239.39 &   255.35& 2,5 \\
 & \nodata  & $\pm$     0.46 & $\pm$     0.00 & $\pm$     4.48 & $\pm$     0.00 & $\pm$     4.22 & $\pm$     7.27 & $\pm$     7.29& \\
GG Tau                         &     6.55 &    14.88 &    49.44 &    94.85 &   193.02 &   547.13 &   757.71 &   764.07& 1,5 \\
 & $\pm$     0.80 & $\pm$     0.80 & $\pm$     2.27 & $\pm$     4.70 & $\pm$    11.90 & $\pm$    27.71 & $\pm$    25.12 & $\pm$    30.25& \\
GM Aur                         &     6.93 &    21.28 &    58.29 &    95.61 & \nodata  &   299.56 &   379.40 &   322.06& 1,5 \\
 & $\pm$     0.00 & $\pm$     0.00 & $\pm$     0.00 & $\pm$     0.00 & \nodata  & $\pm$     9.93 & $\pm$     9.08 & $\pm$     7.12& \\
HL Tau                         &     0.49 &     1.49 &     4.95 &    10.12 &    25.21 &    89.72 &   219.74 &   724.32& 1,5 \\
 & $\pm$     0.11 & $\pm$     0.23 & $\pm$     0.72 & $\pm$     1.49 & $\pm$     3.71 & $\pm$     3.47 & $\pm$     9.31 & $\pm$    11.34& \\
IQ Tau                         &     1.59 &     5.20 &    17.00 &    39.34 &   103.57 &   270.70 &   454.46 &   522.79& 1,5 \\
 & $\pm$     0.94 & $\pm$     2.93 & $\pm$     6.24 & $\pm$    15.46 & $\pm$    40.78 & $\pm$     9.72 & $\pm$    14.23 & $\pm$    19.26& \\
IRAS 04016+2610                & \nodata  & \nodata  &     0.03 &     0.06 &     0.18 &     5.98 &    22.24 &   123.62& 1,4 \\
 & \nodata  & \nodata  & $\pm$     0.00 & $\pm$     0.00 & $\pm$     0.00 & $\pm$     1.84 & $\pm$     3.30 & $\pm$     3.41& \\
IRAS 04169+2702                & \nodata  & \nodata  & \nodata  & \nodata  & \nodata  &     0.34 &     3.01 &    21.68& 4 \\
 & \nodata  & \nodata  & \nodata  & \nodata  & \nodata  & $\pm$     0.17 & $\pm$     0.10 & $\pm$     0.40& \\
IRAS 04248+2612                & \nodata  & \nodata  & \nodata  & \nodata  & \nodata  &    12.73 &    28.83 &    37.09& 4 \\
 & \nodata  & \nodata  & \nodata  & \nodata  & \nodata  & $\pm$     2.12 & $\pm$     4.58 & $\pm$     5.74& \\
IRAS 04302+2247                & \nodata  & \nodata  & \nodata  & \nodata  & \nodata  &     3.33 &    11.76 &    24.88& 4 \\
 & \nodata  & \nodata  & \nodata  & \nodata  & \nodata  & $\pm$     0.00 & $\pm$     0.00 & $\pm$     0.00& \\
IRAS 04325+2402                & \nodata  & \nodata  & \nodata  & \nodata  & \nodata  & \nodata  &     6.34 &    25.82& 4 \\
 & \nodata  & \nodata  & \nodata  & \nodata  & \nodata  & \nodata  & $\pm$     0.00 & $\pm$     0.00& \\
IRAS 04361+2547                & \nodata  & \nodata  & \nodata  & \nodata  & \nodata  &     0.61 &     7.28 &    40.28& 4 \\
 & \nodata  & \nodata  & \nodata  & \nodata  & \nodata  & $\pm$     0.13 & $\pm$     0.96 & $\pm$     2.96& \\
IRAS 04365+2535                & \nodata  & \nodata  & \nodata  & \nodata  & \nodata  &     0.20 &     4.64 &    38.51& 4 \\
 & \nodata  & \nodata  & \nodata  & \nodata  & \nodata  & $\pm$     0.05 & $\pm$     1.10 & $\pm$     7.98& \\
IRAS 04368+2557                & \nodata  & \nodata  & \nodata  & \nodata  & \nodata  &     5.05 &    31.09 &    75.39& 5 \\
 & \nodata  & \nodata  & \nodata  & \nodata  & \nodata  & $\pm$     2.52 & $\pm$    15.55 & $\pm$    37.69& \\
L1551 IRS5                     & \nodata  & \nodata  & \nodata  & \nodata  & \nodata  &    10.88 &    47.06 &   133.53& 4 \\
 & \nodata  & \nodata  & \nodata  & \nodata  & \nodata  & $\pm$     5.48 & $\pm$    23.21 & $\pm$    27.67& \\
LkCa 15                        &     2.75 &    13.59 &    43.87 &    73.39 &   136.38 &   270.45 &   375.58 &   357.05& 1,5 \\
 & $\pm$     0.97 & $\pm$     4.73 & $\pm$    15.13 & $\pm$    25.57 & $\pm$    48.12 & $\pm$     5.73 & $\pm$    10.03 & $\pm$    10.52& \\
RY Tau                         &    42.79 &   146.36 &   345.54 &   507.61 &   891.41 &  2255.74 &  3797.52 &  4746.03& 1,5 \\
 & $\pm$    18.92 & $\pm$    64.16 & $\pm$   150.72 & $\pm$   221.69 & $\pm$   390.17 & $\pm$    66.47 & $\pm$   220.31 & $\pm$   122.37& \\
UY Aur                         &     3.88 &     9.26 &    27.51 &    57.55 &   141.73 &   354.89 &   652.68 &   866.82& 1,5 \\
 & $\pm$     1.83 & $\pm$     4.35 & $\pm$    12.86 & $\pm$    27.20 & $\pm$    68.36 & $\pm$    10.46 & $\pm$    13.82 & $\pm$    25.54& 
\enddata
\tablerefs{1: UBVRI data from \citet{kh95}, 2: BVRI data from the USNO B-1.0 catalog, 3: average BVRI data from the USNO B-1.0 catalog and the NOMAD catalog, 4: JHK data from \citet{kh95}, and 5: JHK data from the 2MASS all-sky survey (the fluxes for DG Tau, DG Tau B, and IRAS 04368+2557 were measured using aperture photometry)}
\tablecomments{Fluxes are in mJy. Values in italics indicate that these are not used when fitting the observed SEDs as higher quality values are available}

\end{deluxetable}

\begin{deluxetable}{lrrrrrrrrr}

\tabletypesize{\tiny}
\tablecolumns{10}
\tablewidth{0pt}
\tablecaption{Mid-IR data for the 30 Taurus-Auriga sources.\label{t:seds_2}}

\tablehead{ Source Name & L & M & N & Q & IRAC 3.6\,$\mu$m & IRAC 4.5\,$\mu$m &
IRAC 5.8\,$\mu$m & IRAC 8.0\,$\mu$m &
References}
%\colhead{(1)} & \colhead{(2)} & \colhead{(3)} & \colhead{(4)} & \colhead{(5)} & \colhead{(6)} & \colhead{(7)} & \colhead{(8)} & \colhead{(9)} & \colhead{(10)}}

%\input{tab8.tex}
\startdata
AA Tau                         & { \it   426.29 } & { \it   297.08 } & { \it   456.14 } & \nodata  &   370.55 &   352.24 &   331.89 &   355.82& 1,2 \\
 & $\pm$  { \it    77.22 } & $\pm$  { \it    43.31 } & $\pm$  { \it   101.69 } & \nodata  & $\pm$    13.64 & $\pm$    12.97 & $\pm$    12.22 & $\pm$     9.83& \\
AB Aur                         & 12758.37 & 11802.62 & { \it 20003.61 } & { \it 57313.72 } & \nodata  & \nodata  & \nodata  & \nodata & 1 \\
 & $\pm$  4022.34 & $\pm$  3097.21 & $\pm$  { \it  4825.28 } & $\pm$  { \it 13825.24 } & \nodata  & \nodata  & \nodata  & \nodata & \\
BP Tau                         & { \it   326.75 } & { \it   298.13 } & { \it   399.26 } & { \it   612.71 } &   318.99 &   281.25 &   231.23 &   335.86& 1,4 \\
 & $\pm$  { \it    74.67 } & $\pm$  { \it    30.05 } & $\pm$  { \it   109.58 } & $\pm$  { \it    61.76 } & $\pm$    12.65 & $\pm$     3.89 & $\pm$     8.33 & $\pm$    16.65& \\
CI Tau                         & { \it   515.01 } & \nodata  & { \it   673.23 } & \nodata  &   640.10 &   558.67 &   487.90 &   572.97& 1,4 \\
 & $\pm$  { \it   141.56 } & \nodata  & $\pm$  { \it   156.72 } & \nodata  & $\pm$     8.08 & $\pm$    34.11 & $\pm$    11.14 & $\pm$    25.35& \\
CoKu Tau 1                     & { \it    23.96 } & \nodata  & \nodata  & \nodata  &    21.72 &    49.55 &   102.98 &   315.60& 1,2 \\
 & $\pm$  { \it     1.76 } & \nodata  & \nodata  & \nodata  & $\pm$     0.80 & $\pm$     2.28 & $\pm$     1.90 & $\pm$     5.81& \\
CY Tau                         & { \it   213.44 } & \nodata  & \nodata  & \nodata  &   203.57 &   170.30 &   148.89 &   135.25& 1,2 \\
 & $\pm$  { \it    24.47 } & \nodata  & \nodata  & \nodata  & $\pm$     5.62 & $\pm$     9.40 & $\pm$     4.11 & $\pm$     2.49& \\
DG Tau                         & { \it  2802.86 } & { \it  4800.51 } & { \it  7062.90 } & { \it 14439.79 } &  1933.80 &  2337.04 &  2697.00 &  3307.00& 1,4 \\
 & $\pm$  { \it   567.96 } & $\pm$  { \it  1152.74 } & $\pm$  { \it  1696.01 } & $\pm$  { \it  3642.47 } & $\pm$    56.63 & $\pm$   153.35 & $\pm$   254.70 & $\pm$   111.03& \\
DG Tau B                       & { \it    93.41 } & { \it     0.00 } & { \it     0.00 } & { \it     0.00 } &    90.62 &   287.88 &   571.47 &   830.28& 1,2 \\
 & $\pm$  { \it     0.00 } & $\pm$  { \it     0.00 } & $\pm$  { \it     0.00 } & $\pm$  { \it     0.00 } & $\pm$     5.00 & $\pm$    15.88 & $\pm$    21.04 & $\pm$    22.93& \\
DL Tau                         & { \it   604.38 } & { \it   628.65 } & { \it   621.41 } & { \it   641.59 } &   484.00 &   533.14 &   525.85 &   590.51& 1,2 \\
 & $\pm$  { \it   188.70 } & $\pm$  { \it    63.37 } & $\pm$  { \it   127.81 } & $\pm$  { \it    64.67 } & $\pm$    17.82 & $\pm$    19.63 & $\pm$    14.52 & $\pm$    16.31& \\
DM Tau                         & { \it   105.50 } & \nodata  & { \it   400.98 } & \nodata  &    55.72 &    37.80 &    23.24 &    22.13& 1,4 \\
 & $\pm$  { \it    36.60 } & \nodata  & $\pm$  { \it   126.03 } & \nodata  & $\pm$     0.37 & $\pm$     1.49 & $\pm$     3.52 & $\pm$     2.76& \\
DN Tau                         & { \it   312.43 } & { \it   324.93 } & { \it   261.63 } & \nodata  &   305.50 &   262.55 &   242.66 &   253.06& 1,2 \\
 & $\pm$  { \it    72.93 } & $\pm$  { \it    41.55 } & $\pm$  { \it    42.33 } & \nodata  & $\pm$    14.05 & $\pm$    14.49 & $\pm$     8.93 & $\pm$     6.99& \\
DO Tau                         & { \it   951.62 } & { \it  1276.84 } & { \it  1481.31 } & { \it  2480.41 } &   965.42 &   988.57 &   956.89 &  1026.19& 1,2 \\
 & $\pm$  { \it   180.88 } & $\pm$  { \it   120.54 } & $\pm$  { \it   258.38 } & $\pm$  { \it   204.91 } & $\pm$    26.66 & $\pm$    45.48 & $\pm$    26.43 & $\pm$    28.34& \\
DR Tau                         & { \it  2315.37 } & { \it  2839.78 } & { \it  3098.98 } & \nodata  &  1858.72 &  1890.13 &  2004.28 &  1736.27& 1,4 \\
 & $\pm$  { \it   866.54 } & $\pm$  { \it  1150.99 } & $\pm$  { \it  1115.34 } & \nodata  & $\pm$   204.37 & $\pm$   151.04 & $\pm$   304.25 & $\pm$   216.95& \\
FT Tau                         & { \it   202.49 } & \nodata  & \nodata  & \nodata  &   249.53 &   248.86 &   205.62 &   271.85& 1,4 \\
 & $\pm$  { \it     0.00 } & \nodata  & \nodata  & \nodata  & $\pm$     2.62 & $\pm$     2.81 & $\pm$     1.69 & $\pm$    10.47& \\
GG Tau                         & { \it   746.69 } & { \it   722.98 } & { \it   945.53 } & { \it   420.42 } &   671.66 &   567.00 &   472.32 &   559.22& 1,4 \\
 & $\pm$  { \it   130.07 } & $\pm$  { \it    83.66 } & $\pm$  { \it   248.12 } & $\pm$  { \it    46.18 } & $\pm$    33.22 & $\pm$    10.93 & $\pm$    18.54 & $\pm$    24.92& \\
GM Aur                         & { \it   161.34 } & \nodata  & { \it   461.93 } & \nodata  &   171.63 &   126.87 &    95.66 &   102.42& 1,3 \\
 & $\pm$  { \it    25.08 } & \nodata  & $\pm$  { \it    67.34 } & \nodata  & $\pm$     2.21 & $\pm$     0.70 & $\pm$     0.79 & $\pm$     2.36& \\
HL Tau                         & { \it  1814.08 } & { \it  4067.90 } & { \it  6150.08 } & { \it 21814.62 } &  3142.80 &  4285.42 &  5119.98 &  4476.30& 1,4 \\
 & $\pm$  { \it   623.29 } & $\pm$  { \it  1501.81 } & $\pm$  { \it  2773.44 } & $\pm$  { \it 10458.28 } & $\pm$   232.28 & $\pm$    41.84 & $\pm$   306.98 & $\pm$    47.81& \\
IQ Tau                         & { \it   412.46 } & \nodata  & { \it   403.47 } & \nodata  &   485.54 &   450.15 &   384.07 &   377.29& 1,2 \\
 & $\pm$  { \it    82.62 } & \nodata  & $\pm$  { \it    65.98 } & \nodata  & $\pm$     8.94 & $\pm$     3.32 & $\pm$     8.84 & $\pm$     2.78& \\
IRAS 04016+2610                & { \it   477.05 } & { \it  2149.58 } & { \it  2805.75 } & { \it 11727.33 } &   906.78 &  1453.51 &  1708.25 &  2634.31& 1,4 \\
 & $\pm$  { \it    13.18 } & $\pm$  { \it    59.37 } & $\pm$  { \it    77.50 } & $\pm$  { \it   323.91 } & $\pm$    69.01 & $\pm$    66.57 & $\pm$    95.73 & $\pm$   133.20& \\
IRAS 04169+2702                & { \it    70.27 } & { \it   227.12 } & { \it   564.88 } & \nodata  &   195.72 &   300.10 &   333.10 &   417.53& 1,4 \\
 & $\pm$  { \it     2.89 } & $\pm$  { \it     4.18 } & $\pm$  { \it    10.40 } & \nodata  & $\pm$     9.27 & $\pm$     9.99 & $\pm$     3.49 & $\pm$    14.07& \\
IRAS 04248+2612                & { \it    28.44 } & \nodata  & { \it   256.55 } & { \it   969.00 } &    65.22 &    67.36 &    78.44 &   109.88& 1,2 \\
 & $\pm$  { \it     8.99 } & \nodata  & $\pm$  { \it    39.68 } & $\pm$  { \it   149.88 } & $\pm$     5.98 & $\pm$     6.18 & $\pm$     7.19 & $\pm$    10.08& \\
IRAS 04302+2247                & { \it    21.40 } & \nodata  & \nodata  & \nodata  &    46.25 &    39.35 &    27.43 &    12.68& 1,4 \\
 & $\pm$  { \it     0.00 } & \nodata  & \nodata  & \nodata  & $\pm$     0.76 & $\pm$     0.46 & $\pm$     0.03 & $\pm$     1.17& \\
IRAS 04325+2402                & \nodata  & \nodata  & \nodata  & \nodata  &    74.88 &    77.34 &    58.96 &    38.81& 2 \\
 & \nodata  & \nodata  & \nodata  & \nodata  & $\pm$     6.87 & $\pm$     7.09 & $\pm$     5.41 & $\pm$     3.56& \\
IRAS 04361+2547                & { \it   103.71 } & { \it   234.08 } & { \it  1059.40 } & \nodata  &   264.48 &   353.50 &   415.47 &   880.85& 1,4 \\
 & $\pm$  { \it    28.49 } & $\pm$  { \it    17.20 } & $\pm$  { \it    77.85 } & \nodata  & $\pm$    24.26 & $\pm$    32.42 & $\pm$    38.10 & $\pm$    80.79& \\
IRAS 04365+2535                & { \it   139.33 } & { \it   276.58 } & { \it   650.91 } & { \it  5280.44 } &   374.30 &   892.98 &  1421.88 &  1469.69& 1,2 \\
 & $\pm$  { \it    28.87 } & $\pm$  { \it    57.30 } & $\pm$  { \it   134.86 } & $\pm$  { \it  1094.05 } & $\pm$    20.65 & $\pm$    32.88 & $\pm$    39.27 & $\pm$    40.59& \\
IRAS 04368+2557                & \nodata  & \nodata  & \nodata  & \nodata  &   166.87 &   221.00 &   264.57 &    82.58& 2 \\
 & \nodata  & \nodata  & \nodata  & \nodata  & $\pm$    15.30 & $\pm$    20.27 & $\pm$    24.27 & $\pm$     7.57& \\
L1551 IRS5                     & { \it   447.41 } & { \it  1422.04 } & { \it  5266.47 } & { \it 25042.07 } &   598.89 &  1375.48 &  3105.60 &  4908.54& 1,4 \\
 & $\pm$  { \it   179.84 } & $\pm$  { \it   814.78 } & $\pm$  { \it  1091.15 } & $\pm$  { \it  5188.44 } & $\pm$     5.83 & $\pm$     2.70 & $\pm$    34.90 & $\pm$    41.39& \\
LkCa 15                        & { \it   253.14 } & \nodata  & { \it   201.46 } & \nodata  &   267.79 &   207.04 &   142.68 &   163.29& 1,4 \\
 & $\pm$  { \it    16.77 } & \nodata  & $\pm$  { \it     7.42 } & \nodata  & $\pm$     1.53 & $\pm$     2.62 & $\pm$     3.77 & $\pm$     1.06& \\
RY Tau                         & { \it  5677.26 } & { \it  7554.89 } & { \it 15922.10 } & \nodata  &  5749.96 &  5298.11 &  4403.50 &  5829.60& 1,4 \\
 & $\pm$  { \it  1057.69 } & $\pm$  { \it  2110.88 } & $\pm$  { \it  5629.73 } & \nodata  & $\pm$    72.53 & $\pm$    57.59 & $\pm$   157.21 & $\pm$   161.56& \\
UY Aur                         & { \it  1398.63 } & \nodata  & { \it  2809.14 } & { \it  4939.96 } &  1072.04 & \nodata  &  1106.39 &  1713.42& 1,3 \\
 & $\pm$  { \it   253.34 } & \nodata  & $\pm$  { \it   409.52 } & $\pm$  { \it   720.16 } & $\pm$    17.77 & \nodata  & $\pm$     8.15 & $\pm$     4.73& 
\enddata
\tablerefs{1: LMNQ data from \citet{kh95}, 2: IRAC data from \citet{hartmann05}, 3: IRAC data from \citet{luhman06}, 4: IRAC data measured using data retrieved from the Spitzer Space Telescope Archive}
\tablecomments{Fluxes are in mJy. Values in italics indicate that these are not used when fitting the observed SEDs as higher quality values are available}

\end{deluxetable}

\begin{deluxetable}{lrrrrrrr}

\tabletypesize{\tiny}
\tablecolumns{8}
\tablewidth{0pt}
\tablecaption{Far-IR data for the 30 Taurus-Auriga sources.\label{t:seds_3}}

\tablehead{ Source Name & MIPS~24 & MIPS~70 & IRAS~12 & IRAS~25 & IRAS~60 & IRAS~100 & References}
%\\\colhead{(1)} & \colhead{(2)} & \colhead{(3)} & \colhead{(4)} & \colhead{(5)} & \colhead{(6)} & \colhead{(7)} & \colhead{(8)}}

%\input{tab9.tex}
\startdata
AA Tau                         &   502.83 &   950.05 &   430.00 & { \it   610.00 } & { \it  1230.00 } & { \it  3290.00 }& 1,2 \\
 & $\pm$     2.80 & $\pm$    95.00 & $\pm$    33.00 & $\pm$ { \it    32.00 } & $\pm$ { \it    96.00 } & $\pm$ { \it   601.00 }& \\
AB Aur                         & \nodata  & \nodata  & 28950.00 & 49780.00 & 115680.00 & 114470.00& 2 \\
 & \nodata  & \nodata  & $\pm$    63.00 & $\pm$   107.00 & $\pm$    72.00 & $\pm$  1956.00& \\
BP Tau                         & \nodata  & \nodata  &   450.00 &   590.00 &   440.00 &   920.00& 2 \\
 & \nodata  & \nodata  & $\pm$    35.00 & $\pm$    36.00 & $\pm$     2.00 & $\pm$   226.00& \\
CI Tau                         &   956.95 & \nodata  &   780.00 & { \it  1300.00 } &  2150.00 & $<$  2540.00& 1,2,3 \\
 & $\pm$     5.70 & \nodata  & $\pm$    25.00 & $\pm$ { \it    46.00 } & $\pm$    71.00 & \nodata& \\
CoKu Tau 1                     &  3453.61 & \nodata  &  1180.00 & { \it  2740.00 } & $<$  7970.00 & $<$ 71090.00& 1,2,3 \\
 & $\pm$    38.10 & \nodata  & $\pm$    26.00 & $\pm$ { \it    63.00 } & \nodata & \nodata& \\
CY Tau                         &   130.48 &    98.86 &   190.00 & { \it   270.00 } & { \it   140.00 } & \nodata & 1,2 \\
 & $\pm$     1.38 & $\pm$     9.89 & $\pm$    45.00 & $\pm$ { \it    35.00 } & $\pm$ { \it    90.00 } & \nodata & \\
DG Tau                         &  9332.72 &  9824.49 & \nodata  & \nodata  & \nodata  & \nodata & 1 \\
 & $\pm$   854.50 & $\pm$   982.45 & \nodata  & \nodata  & \nodata  & \nodata & \\
DG Tau B                       &  4641.72 &  7836.02 & \nodata  & \nodata  & \nodata  & \nodata & 1 \\
 & $\pm$    99.20 & $\pm$   783.60 & \nodata  & \nodata  & \nodata  & \nodata & \\
DL Tau                         &   906.17 &   888.20 &   970.00 & { \it  1320.00 } & { \it  1390.00 } & { \it  2830.00 }& 1,2 \\
 & $\pm$     4.80 & $\pm$    88.82 & $\pm$    34.00 & $\pm$ { \it    45.00 } & $\pm$ { \it    84.00 } & $\pm$ { \it   949.00 }& \\
DM Tau                         & \nodata  & \nodata  & $<$   270.00 &   350.00 &   830.00 & $<$  7210.00& 2,3 \\
 & \nodata  & \nodata  & \nodata & $\pm$    33.00 & $\pm$   104.00 & \nodata& \\
DN Tau                         &   410.03 &   423.01 &   350.00 & { \it   600.00 } & { \it   650.00 } & { \it $<$  5787.00 }& 1,2 \\
 & $\pm$     2.60 & $\pm$    42.30 & $\pm$    30.00 & $\pm$ { \it    53.00 } & $\pm$ { \it    93.00 } & \nodata& \\
DO Tau                         &  3115.98 &  2745.71 &  1880.00 & { \it  4070.00 } & { \it  6330.00 } & { \it  8550.00 }& 1,2 \\
 & $\pm$    32.30 & $\pm$   274.57 & $\pm$    34.00 & $\pm$ { \it    41.00 } & $\pm$ { \it   106.00 } & $\pm$ { \it  1483.00 }& \\
DR Tau                         & \nodata  & \nodata  &  3160.00 &  4300.00 &  5510.00 &  6980.00& 2 \\
 & \nodata  & \nodata  & $\pm$    29.00 & $\pm$    51.00 & $\pm$    44.00 & $\pm$  1144.00& \\
FT Tau                         & \nodata  & \nodata  &   360.00 &   570.00 &   820.00 &  1800.00& 2 \\
 & \nodata  & \nodata  & $\pm$    45.00 & $\pm$    37.00 & $\pm$    41.00 & $\pm$  1287.00& \\
GG Tau                         & \nodata  & \nodata  &  1200.00 &  1720.00 &  3120.00 &  5480.00& 2 \\
 & \nodata  & \nodata  & $\pm$    33.00 & $\pm$    39.00 & $\pm$    71.00 & $\pm$   365.00& \\
GM Aur                         &   747.62 &  1877.76 &   250.00 & { \it  1070.00 } & { \it  3080.00 } & { \it  3440.00 }& 1,2 \\
 & $\pm$     6.55 & $\pm$    17.80 & $\pm$    31.00 & $\pm$ { \it    42.00 } & $\pm$ { \it   112.00 } & $\pm$ { \it  1969.00 }& \\
HL Tau                         & \nodata  & \nodata  &  9740.00 & 31180.00 & 76260.00 & 77950.00& 2 \\
 & \nodata  & \nodata  & $\pm$    31.00 & $\pm$    73.00 & $\pm$   810.00 & $\pm$  1394.00& \\
IQ Tau                         &   500.09 &   579.59 &   500.00 & { \it   650.00 } & { \it   770.00 } & { \it $<$  1800.00 }& 1,2,3 \\
 & $\pm$     3.00 & $\pm$    57.96 & $\pm$    29.00 & $\pm$ { \it    30.00 } & $\pm$ { \it    43.00 } & \nodata& \\
IRAS 04016+2610                & 11992.36 & \nodata  &  3640.00 & { \it 15810.00 } & 48790.00 & 55690.00& 1,2 \\
 & $\pm$   257.11 & \nodata  & $\pm$   109.20 & $\pm$ { \it   632.40 } & $\pm$  3903.20 & $\pm$  7239.70& \\
IRAS 04169+2702                &  4519.14 & 11745.45 &   750.00 & { \it  5210.00 } & { \it 17000.00 } & { \it 17460.00 }& 1,2 \\
 & $\pm$   821.84 & $\pm$  1174.55 & $\pm$    52.50 & $\pm$ { \it   312.60 } & $\pm$ { \it  1700.00 } & $\pm$ { \it  2269.80 }& \\
IRAS 04248+2612                &   835.33 &  3640.03 & $<$   360.00 & { \it  1330.00 } & { \it  4620.00 } & { \it  9260.00 }& 1,2 \\
 & $\pm$     4.80 & $\pm$   364.00 & \nodata & $\pm$ { \it   106.40 } & $\pm$ { \it   415.80 } & $\pm$ { \it   833.40 }& \\
IRAS 04302+2247                &   241.09 &  4774.75 & $<$   250.00 & { \it   440.00 } & { \it  6400.00 } & { \it  9430.00 }& 1,2 \\
 & $\pm$     1.80 & $\pm$   477.48 & \nodata & $\pm$ { \it    83.60 } & $\pm$ { \it   640.00 } & $\pm$ { \it  1131.60 }& \\
IRAS 04325+2402                &  1860.53 &  8042.05 & $<$   250.00 & { \it  2100.00 } & { \it 12860.00 } & { \it 22350.00 }& 1,2 \\
 & $\pm$    11.60 & $\pm$   804.21 & \nodata & $\pm$ { \it   168.00 } & $\pm$ { \it  1157.40 } & $\pm$ { \it  3576.00 }& \\
IRAS 04361+2547                & \nodata  & \nodata  &  1820.00 & 18870.00 & 44750.00 & 35430.00& 2 \\
 & \nodata  & \nodata  & $\pm$   109.20 & $\pm$  1132.20 & $\pm$  5370.00 & $\pm$  4251.60& \\
IRAS 04365+2535                &  6118.02 & \nodata  &  1190.00 & { \it  8620.00 } & 36010.00 & 39250.00& 1,2 \\
 & $\pm$   399.90 & \nodata  & $\pm$   142.80 & $\pm$ { \it   517.20 } & $\pm$  4321.20 & $\pm$  5887.50& \\
IRAS 04368+2557                &   542.92 & \nodata  & $<$   250.00 & { \it   740.00 } & 17770.00 & 73260.00& 1,2 \\
 & $\pm$     3.20 & \nodata  & \nodata & $\pm$ { \it    66.60 } & $\pm$  1599.30 & $\pm$ 11721.60& \\
L1551 IRS5                     & \nodata  & \nodata  & 10040.00 & 106200.00 & 372900.00 & 457900.00& 2 \\
 & \nodata  & \nodata  & $\pm$   502.00 & $\pm$  4248.00 & $\pm$ 18645.00 & $\pm$ 59527.00& \\
LkCa 15                        & \nodata  & \nodata  &   270.00 &   390.00 &  1500.00 &  1610.00& 2 \\
 & \nodata  & \nodata  & $\pm$    30.00 & $\pm$    33.00 & $\pm$    48.00 & $\pm$   200.00& \\
RY Tau                         & 17862.32 &  9632.66 & 17740.00 & { \it 26480.00 } & { \it 18910.00 } & { \it 13500.00 }& 1,2 \\
 & $\pm$  4424.47 & $\pm$   963.27 & $\pm$    27.00 & $\pm$ { \it    54.00 } & $\pm$ { \it    66.00 } & $\pm$ { \it  2501.00 }& \\
UY Aur                         & \nodata  & \nodata  &  3710.00 &  6870.00 &  7580.00 &  9400.00& 2 \\
 & \nodata  & \nodata  & $\pm$    43.00 & $\pm$    39.00 & $\pm$    87.00 & $\pm$   698.00& 
\enddata
\tablerefs{1: MIPS data measured using data retrieved from the Spitzer Space Telescope Archive, 2: IRAS data from \citet{weaver92}, 3: IRAS data from IRAS point-source catalog}
\tablecomments{Fluxes are in mJy. Values in italics indicate that these are not used when fitting the observed SEDs as higher quality values are available}

\end{deluxetable}

\begin{deluxetable}{lrrrrrr}

\tabletypesize{\tiny}
\tablecolumns{7}
\tablewidth{0pt}
\tablecaption{Sub-mm data for the 30 Taurus-Auriga sources.\label{t:seds_4}}

\tablehead{ Source Name & SHARC 350\,\microns & SCUBA 450\,\microns & SCUBA 850\,\microns & CSO 624\,\microns & CSO 729\,\microns & References}
% \\\colhead{(1)} & \colhead{(2)} & \colhead{(3)} & \colhead{(4)} & \colhead{(5)} & \colhead{(6)}}

%\input{tab10.tex}
\startdata
AA Tau                         &   825.00 &   415.00 &   144.00 & \nodata  &   310.00& 1,2 \\
 & $\pm$    50.00 & $\pm$    84.00 & $\pm$     5.00 & \nodata  & $\pm$    60.00& \\
AB Aur                         &  8930.00 &  3820.00 &   359.00 & \nodata  & \nodata & 1 \\
 & $\pm$  1410.00 & $\pm$   570.00 & $\pm$    67.00 & \nodata  & \nodata & \\
BP Tau                         & \nodata  & $<$   456.00 &   130.00 & \nodata  & \nodata & 1 \\
 & \nodata  & \nodata & $\pm$     7.00 & \nodata  & \nodata & \\
CI Tau                         &  1725.00 &   846.00 &   324.00 &  1300.00 &   850.00& 1,2 \\
 & $\pm$    55.00 & $\pm$    89.00 & $\pm$     6.00 & $\pm$   210.00 & $\pm$   150.00& \\
CoKu Tau 1                     & \nodata  & $<$   522.00 &    35.00 & \nodata  & \nodata & 1 \\
 & \nodata  & \nodata & $\pm$     7.00 & \nodata  & \nodata & \\
CY Tau                         & $<$  1839.00 & $<$   210.00 &   140.00 & \nodata  &   240.00& 1,2 \\
 & \nodata & \nodata & $\pm$     5.00 & \nodata  & $\pm$    40.00& \\
DG Tau                         & $<$  5173.00 & $<$  3950.00 & $<$  1100.00 & $<$  1210.00 & $<$   860.00& 1,2 \\
 & \nodata & \nodata & \nodata & \nodata & \nodata& \\
DG Tau B                       & $<$  5173.00 & $<$  3950.00 & $<$  1100.00 & $<$  1210.00 & $<$   860.00& 1,2 \\
 & \nodata & \nodata & \nodata & \nodata & \nodata& \\
DL Tau                         &  1390.00 &  1280.00 &   440.00 &   880.00 &   530.00& 1,2 \\
 & $\pm$   180.00 & $\pm$   170.00 & $\pm$    40.00 & $\pm$   140.00 & $\pm$    90.00& \\
DM Tau                         &  1077.00 & \nodata  &   237.00 &   390.00 &   470.00& 1,2 \\
 & $\pm$    49.00 & \nodata  & $\pm$    12.00 & $\pm$   130.00 & $\pm$    80.00& \\
DN Tau                         &   615.00 & $<$   703.00 &   201.00 & \nodata  &   380.00& 1,2 \\
 & $\pm$    64.00 & \nodata & $\pm$     7.00 & \nodata  & $\pm$    80.00& \\
DO Tau                         & \nodata  &   734.00 &   258.00 &   700.00 &   510.00& 1,2 \\
 & \nodata  & $\pm$    50.00 & $\pm$    42.00 & $\pm$   100.00 & $\pm$   100.00& \\
DR Tau                         & \nodata  &  2380.00 &   533.00 & \nodata  &   400.00& 1,2 \\
 & \nodata  & $\pm$   172.00 & $\pm$     7.00 & \nodata  & $\pm$    80.00& \\
FT Tau                         &  1106.00 &   437.00 &   121.00 &   260.00 &   250.00& 1,2 \\
 & $\pm$    82.00 & $\pm$    56.00 & $\pm$     5.00 & $\pm$   100.00 & $\pm$    50.00& \\
GG Tau                         &  6528.00 &  2726.00 &  1255.00 &  1370.00 &  1250.00& 1,2 \\
 & $\pm$   153.00 & $\pm$   250.00 & $\pm$    57.00 & $\pm$   170.00 & $\pm$     8.00& \\
GM Aur                         &  3419.00 & \nodata  & \nodata  &  1340.00 &   850.00& 1,2 \\
 & $\pm$   133.00 & \nodata  & \nodata  & $\pm$   330.00 & $\pm$    90.00& \\
HL Tau                         & 23888.00 & 10400.00 &  2360.00 &  5450.00 &  3200.00& 1,2 \\
 & $\pm$   149.00 & $\pm$  1400.00 & $\pm$    90.00 & $\pm$   290.00 & $\pm$   100.00& \\
IQ Tau                         & \nodata  &   425.00 &   178.00 & \nodata  & \nodata & 1 \\
 & \nodata  & $\pm$    26.00 & $\pm$     3.00 & \nodata  & \nodata & \\
IRAS 04016+2610                & 12477.00 & \nodata  & \nodata  & \nodata  & \nodata & 1 \\
 & $\pm$   193.00 & \nodata  & \nodata  & \nodata  & \nodata & \\
IRAS 04169+2702                &  7344.00 & \nodata  & \nodata  & \nodata  & \nodata & 1 \\
 & $\pm$   152.00 & \nodata  & \nodata  & \nodata  & \nodata & \\
IRAS 04248+2612                &  1178.00 & \nodata  & \nodata  & \nodata  & \nodata & 1 \\
 & $\pm$    30.00 & \nodata  & \nodata  & \nodata  & \nodata & \\
IRAS 04302+2247                &  2869.00 & \nodata  & \nodata  & \nodata  & \nodata & 1 \\
 & $\pm$    21.00 & \nodata  & \nodata  & \nodata  & \nodata & \\
IRAS 04325+2402                & \nodata  &   606.00 &   186.00 & \nodata  & \nodata & 1 \\
 & \nodata  & $\pm$   185.00 & $\pm$    11.00 & \nodata  & \nodata & \\
IRAS 04361+2547                & \nodata  &  1302.00 &   275.00 & \nodata  & \nodata & 1 \\
 & \nodata  & $\pm$   168.00 & $\pm$     8.00 & \nodata  & \nodata & \\
IRAS 04365+2535                & \nodata  &  2928.00 &   622.00 & \nodata  & \nodata & 1 \\
 & \nodata  & $\pm$   230.00 & $\pm$    13.00 & \nodata  & \nodata & \\
IRAS 04368+2557                & \nodata  &  2849.00 &   895.00 & \nodata  & \nodata & 1 \\
 & \nodata  & $\pm$   222.00 & $\pm$    11.00 & \nodata  & \nodata & \\
L1551 IRS5                     & 100423.00 & \nodata  & \nodata  & \nodata  & \nodata & 1 \\
 & $\pm$   812.00 & \nodata  & \nodata  & \nodata  & \nodata & \\
LkCa 15                        &  1235.00 & \nodata  &   428.00 & \nodata  & \nodata & 1 \\
 & $\pm$    80.00 & \nodata  & $\pm$    11.00 & \nodata  & \nodata & \\
RY Tau                         &  2439.00 &  1920.00 &   560.00 & \nodata  & \nodata & 1 \\
 & $\pm$   330.00 & $\pm$   160.00 & $\pm$    30.00 & \nodata  & \nodata & \\
UY Aur                         &   542.00 & $<$   523.00 &   102.00 & \nodata  & \nodata & 1 \\
 & $\pm$    77.00 & \nodata & $\pm$     6.00 & \nodata  & \nodata & 
\enddata
\tablerefs{1: SHARC~II and SCUBA data from the compilation presented in \citet{andrews05}, 2: CSO data from \cite{beckwith91}}
\tablecomments{Fluxes are in mJy. Values in italics indicate that these are not used when fitting the observed SEDs as higher quality values are available}

\end{deluxetable}

\begin{deluxetable}{lccccccccc}

\tabletypesize{\tiny}
\tablecolumns{10}
\tablewidth{0pt}
\tablecaption{Apertures assumed for the SED fitting.\label{t:seds_ap}}

\tablehead{Source Name & UBVRI & JHK & LM & IRAC & MIPS 24\microns & MIPS 70\microns & IRAS 12 \& 25\microns & IRAS 60 \& 100\microns & sub-mm \\
\colhead{} & \colhead{\arcsec} & \colhead{\arcsec} & \colhead{\arcsec} & \colhead{\arcsec} & \colhead{\arcsec} & \colhead{\arcsec} & \colhead{\arcsec} & \colhead{\arcsec} & \colhead{\arcsec}}

\startdata
AA Tau                         &      5  &	3  & \nodata &      5  &     10  &     20  &	 60  & \nodata &     30   \\
AB Aur                         &      5  &	3  &	 15  & \nodata & \nodata & \nodata &	 60  &    120  &     30   \\
BP Tau                         &      5  &	3  & \nodata &      5  & \nodata & \nodata &	 60  &    120  &     30   \\
CI Tau                         &      5  &	3  & \nodata &      5  &     10  & \nodata &	 60  &    120  &     30   \\
CoKu~Tau 1                     &      5  &	3  & \nodata &      5  &     10  & \nodata &	 60  &    120  &     30   \\
CY Tau                         &      5  &	3  & \nodata &      5  &     10  &     20  &	 60  & \nodata &     30   \\
DG Tau                         &      5  &     10  & \nodata &      5  &     10  &     20  & \nodata & \nodata &     30   \\
DG Tau B                       & \nodata &     10  & \nodata &      5  &     10  &     20  & \nodata & \nodata &     30   \\
DL Tau                         &      5  &	3  & \nodata &      5  &     10  &     20  &	 60  & \nodata &     30   \\
DM Tau                         &      5  &	3  & \nodata &      5  & \nodata & \nodata &	 60  &    120  &     30   \\
DN Tau                         &      5  &	3  & \nodata &      5  &     10  &     20  &	 60  & \nodata &     30   \\
DO Tau                         &      5  &	3  & \nodata &      5  &     10  &     20  &	 60  & \nodata &     30   \\
DR Tau                         &      5  &	3  & \nodata &      5  & \nodata & \nodata &	 60  &    120  &     30   \\
FT Tau                         &      5  &	3  & \nodata &      5  & \nodata & \nodata &	 60  &    120  &     30   \\
GG Tau                         &      5  &	3  & \nodata &      5  & \nodata & \nodata &	 60  &    120  &     30   \\
GM Aur                         &      5  &	3  & \nodata &      5  &     10  &     20  &	 60  & \nodata &     30   \\
HL Tau                         &      5  &	3  & \nodata &      5  & \nodata & \nodata &	 60  &    120  &     30   \\
IQ Tau                         &      5  &	3  & \nodata &      5  &     10  &     20  &	 60  & \nodata &     30   \\
IRAS 04016+2610                &      5  &     15  & \nodata &      5  &     10  & \nodata &	 60  &    120  &     30   \\
IRAS 04169+2702                & \nodata &     15  & \nodata &      5  &     10  &     20  &	 60  & \nodata &     30   \\
IRAS 04248+2612                & \nodata &     15  & \nodata &     26  &     10  &     20  &	 60  & \nodata &     30   \\
IRAS 04302+2247                & \nodata &     15  & \nodata &     26  &     10  &     20  &	 60  & \nodata &     30   \\
IRAS 04325+2402                & \nodata &     15  & \nodata &     26  &     10  &     20  &	 60  & \nodata &     30   \\
IRAS 04361+2547                & \nodata &     15  & \nodata &     35  & \nodata & \nodata &	 60  &    120  &     30   \\
IRAS 04365+2535                & \nodata &     15  & \nodata &      5  &     10  & \nodata &	 60  &    120  &     30   \\
IRAS 04368+2557                & \nodata &    100  & \nodata &    100  &     10  & \nodata &	 60  &    120  &     30   \\
L1551 IRS5                     & \nodata &     15  & \nodata &     15  & \nodata & \nodata &	 60  &    120  &     30   \\
LkCa 15                        &      5  &	3  & \nodata &     13  & \nodata & \nodata &	 60  &    120  &     30   \\
RY Tau                         &      5  &	3  & \nodata &      5  &     10  &     20  &	 60  & \nodata &     30   \\
UY Aur                         &      5  &	3  & \nodata &      5  & \nodata & \nodata &	 60  &    120  &     30   \\
\enddata
\tablecomments{The apertures quoted are in arcseconds. When PSF photometry is done, we took the aperture to be slightly larger than the FWHM of the PSF. The apertures for the UBVRIJHKLM data from \citet{kh95} are estimates. The 3\arcsec aperture for the 2MASS data is slightly larger than the 2\arcsec pixel size of 2MASS data. The JHK apertures for DG Tau, DG Tau B and IRAS 04368+2557 are those used to carry out aperture photometry on 2MASS data. An aperture of 5\arcsec was used for the IRAC data from \citet{hartmann05} and \citet{luhman06} (the latter used PSF photometry, and 5\arcsec is likely an upper limit on the source sizes). The apertures for the IRAC data are those used for aperture photometry by \citet{luhman06} or by the authors of this paper. The MIPS PSF full width half maxima are 6 and 18\arcsec for MIPS 24 and 70\,\microns respectively. Therefore, we used apertures of 10 and 20\arcsec respectively. The apertures on the IRAS telescope were rectangular, therefore we are only able to use estimates of an effective aperture radius. We used 60\arcsec for the 12 and 25\,\microns data, and 120\arcsec for the 60 and 100\,\microns data, Finally, \citet{andrews05} used a 30\arcsec aperture to measure the SHARC~II 350\,\microns sub-mm fluxes, but we do not have information concerning the SCUBA and CSO data. Since the FWHM of all these instruments is comparable, we assumed a 30\arcsec aperture for all sub-mm fluxes.}

\end{deluxetable}

\end{document}